\documentclass[11pt,A4paper]{article}%
\usepackage{jheppub}
\usepackage{amsmath,amssymb,graphicx,epstopdf,enumerate,amsfonts,booktabs,color,soul}
\usepackage{slashed}
\usepackage{float,placeins}
\usepackage{appendix}
\usepackage{relsize}
\usepackage{xcolor}
\usepackage{lscape}
\usepackage[caption=false]{subfig}
\usepackage{hyperref}
\usepackage{xspace}
\usepackage{bm}
\usepackage{amsmath}
\usepackage{amsfonts}
\usepackage{amssymb}
\usepackage{graphicx}%
\providecommand{\U}[1]{\protect\rule{.1in}{.1in}}
\newcommand\utimes{\mathbin{\ooalign{$\cup$\cr   \hfil\raise0.42ex\hbox{$\scriptscriptstyle\times$}\hfil\cr}}}
\affiliation[a]{Center for Fundamental Physics, School of Physical Science and Technology, ShanghaiTech University, Shanghai, PRC}
\affiliation[b]{Department of Physics, Ateneo de Manila University, Loyola Heights, Quezon City, Philippines}
\emailAdd{zhoujr@shanghaitech.edu.cn}
\emailAdd{holao@ateneo.edu}
\emailAdd{yangyi3@shanghaitech.edu.cn}
\abstract{We present a graphical method for proving holographic entanglement entropy inequalities (HEIs) in general multipartite systems. By introducing a geometric representation of the entanglement structure, we develop a systematic approach that enables one to visualize and verify the validity of HEIs for any number of subsystems $n$. Several theorems are established to formalize this method, and explicit examples are provided for systems with $n = 4$ to $7$ entangled regions.}
\begin{document}

\title{A graphical framework for proving holographic entanglement entropy inequalities in multipartite systems}
\author{$\text{Chia-Jui Chou}^a$, $\text{Hans B. Lao}^b$, $\text{Yi Yang}^a$}
\maketitle

\section{Introduction}

Entanglement is one of the defining features that distinguishes quantum physics from classical physics, and it can be quantified using the von Neumann entropy. Although calculating entanglement entropy in general quantum field theories is often challenging, Ryu and Takayanagi proposed a geometric prescription based on the AdS/CFT correspondence \cite{0603001,0605073}. According to their proposal, the holographic entanglement entropy (HEE) of a boundary region $A$ is determined by the area of the minimal codimension-2 bulk surface $\mathcal{E}_{{A}}$ that is homologous to $A$:
\begin{equation}
S_{{A}}=\min_{X}\frac{\text{Area} \left( X \right)}{4G_{N}^{\left(
d+2\right)  }}\text{, }X=\left\{  \mathcal{E}_{{A}}\Big\vert~\left.
\mathcal{E}_{{A}}\right\vert _{\mathcal{\partial M}}=\partial{A}\text{;
}\exists\mathcal{R}_{{A}}\subset\mathcal{M}\text{, }\partial\mathcal{R}_{{A}%
}=\mathcal{E}_{{A}}\cup{A}\right\}  , \label{RT formula}%
\end{equation}
where $\mathcal{E}_{{A}}$ is the Ryu–Takayanagi (RT) surface anchored on
$\partial{A}$, and $\mathcal{R}%
_{{A}}$ is the corresponding entanglement wedge bounded by $\mathcal{E}_{A}$ and $A$. The surface must satisfy a homology constraint, meaning that $\mathcal{E}_{A}$ can be continuously deformed to $A$ within the bulk region $\mathcal{R}_{{A}}$.

This geometric prescription has allowed one to analyze various properties of entanglement entropy in holographic theories \cite{0606184,0705.0016,1006.0047,1102.0440,1304.4926,1609.01287}. In ordinary quantum systems, the von Neumann entropy obeys several fundamental inequalities, such as subadditivity  (SA) \cite{Araki&Lieb} and strong subadditivity (SSA) \cite{Lieb&Ruskai}, which are central to quantum information theory. However, deriving all such inequalities directly from quantum mechanics is difficult due to the non-commutativity of the reduced density matrices. 

In contrast, within holographic theories, many entropy inequalities can be proven geometrically using the RT prescription. These include the proofs of strong  subadditivity
\cite{0704.3719,1211.3494} and monogamy of mutual information (MMI)
\cite{1211.3494,1107.2940}. Moreover, efforts have been made to classify all possible holographic entropy inequalities systematically, leading to constructions such as the holographic entropy cone \cite{1505.07839} and the $I_{n}$-theorem for primitive information
quantities in multipartite systems \cite{1612.02437,1808.07871}. Some of the inequalities derived holographically may not hold for general quantum systems, but they often reveal deeper insights into the geometric and entanglement structure of holographic states. Such states are sometimes called \textit{geometric states} \cite{1808.07871,1905.06985,2003.03933,2008.12430}.

A related development is the proposed correspondence between the entanglement of purification (EoP) and the minimal entanglement wedge cross section (EWCS) \cite{1708.09393,1709.07424}. The EoP measures the total correlations between two subsystems $A$ and $B$, and has been shown to characterize many-body behaviors,  such as $Z_{2}$ symmetry breaking in the transverse-field Ising model \cite{1802.09545,1902.02369}.

In our previous work \cite{2011.02790}, we introduced notations ${A} _{\left\langle ij
\right\rangle }$ and ${A}_{\left[  ij\right]  }$ to describe multipartite systems, allowing one to express entropy inequalities in a balanced form with an equal number of terms on both sides.

In this paper, we extend these ideas by developing a graphical framework to systematically prove holographic entanglement entropy inequalities in multipartite systems. In particular, we establish the \textit{Compatibility Theorem}, \textit{Gapless Theorem}, \textit{Cut Theorem}, and \textit{Configuration Theorem}, which together provide powerful tools for verifying the validity of Holographic Entropy Inequalities (HEIs), irrespective of whether the entangling regions are
connected or disconnected.

Section~2 introduces HEIs in multipartite systems, and both the $I$-basis and the simplex basis are defined. In Section~3, we review the strong subadditivity (SSA) and the monogamy of mutual information (MMI) in the tripartite case. Section~4 presents the main results of this work, outlining a general strategy for proving superbalanced HEIs in multipartite systems through the establishment of the four central theorems mentioned above. In Section~5, we illustrate the method with explicit examples of HEIs in 6-partite and 7-partite systems. We conclude in Section~6 with a summary and outlook.

\section{Holographic Entanglement Entropy Inequality}

The simplest HEI is the
subadditivity (SA) in a bipartite system,
\begin{equation}
S_{12}\leq S_{1}+S_{2}. \label{SA}%
\end{equation}
For a tripartite system, two fundamental inequalities are known: strong subadditivity (SSA) and monogamy of mutual information (MMI),
\begin{align}
S_{12}+S_{23}  &  \geq S_{123}+S_{2},\label{SSA}\\
S_{12}+S_{23}+S_{13}  &  \geq S_{1}+S_{2}+S_{3}+S_{123}. \label{MMI}%
\end{align}
Both can be proven straightforwardly using the Ryu–Takayanagi (RT) prescription.

However, for systems with more than three subsystems, HEIs become far more complicated. In what follows, we develop a systematic approach to prove such inequalities in general multipartite settings. 


\subsection{$I$-basis}

The inequalities in eqs.(\ref{SA})-(\ref{MMI}) are written in the so-called $S$-basis, which directly involve the entropies of the subsystems. In some situations, however, it is more convenient to work with the $I$-basis, first introduced in \cite{1812.08133}.

For example, we define the mutual information,
\begin{equation}
I_{12}=S_{1}+S_{2}-S_{12}, \label{I12}%
\end{equation}
and the tripartite information,
\begin{equation}
I_{123}=S_{1}+S_{2}+S_{3}-S_{12}-S_{13}-S_{23}+S_{123}. \label{I123}
\end{equation}
In this notation, SA eq.(\ref{SA}) becomes simply $I_{12}\geq0$, while MMI eq.(\ref{MMI}) becomes $I_{123}\leq0$, respectively.

This idea can be generalized to the $n$-partite information
\cite{1812.08133},%
\begin{equation}
I_{k_{1}\cdots k_{m}}=\sum S_{k_{\sigma}}-\sum S_{k_{\sigma_{1}}k_{\sigma_{2}%
}}+\sum S_{k_{\sigma_{1}}k_{\sigma_{2}}k_{\sigma_{3}}}+\cdots+\left(
-1\right)  ^{m-1}S_{k_{1}\cdots k_{m}} \label{In}%
\end{equation}
or more compactly,
\begin{equation}
I_{K}=\sum_{J\subseteq K}\left(  -1\right)  ^{\left\vert J\right\vert +1}%
S_{J},
\end{equation}
where $K$ is a set of indices and $\left\vert J\right\vert $ denotes the number of elements in $J$.

Note that while $I_{12}\geq0$ and $I_{123}\leq0$ have
definite signs, higher-order $I_K$ with $|K|>3$ does not necessarily carry a fixed sign.

Since the RT surface areas diverge near the boundary, an HEI is physically meaningful only if the divergences cancel. This happens when the sets of positive and negative terms share the same indices—such inequalities are called \textit{balanced}. By construction, any $n$-partite information with $\left\vert K\right\vert \geq2$ is balanced. Therefore, any balanced inequality can be expanded in terms of $I_K$ with $\left\vert K\right\vert \geq2$:
\begin{equation}
Q_{n}^{b}=\sum_{\left\vert K\right\vert \geq2}q_{K}I_{K}.
\end{equation}
As shown in \cite{2002.04558}, the fundamental or non-redundant inequalities are in fact superbalanced, meaning that they involve only terms with $\left\vert K\right\vert \geq3$,
\begin{equation}
Q_{n}^{s}=\sum_{\left\vert K\right\vert \geq3}q_{K}I_{K}.
\end{equation}
For example, both SA eq.(\ref{SA}) and MMI eq.(\ref{MMI}) are balanced, but only
MMI is superbalanced.

All holographic inequalities can be written as positive combinations of non-redundant HEIs and  mutual information $I_{ij}$. Hence, proving all non-redundant (superbalanced) HEIs is sufficient to establish the entire set of holographic entropy inequalities that define the holographic
inequality cone.

The $I$-basis is particularly convenient for manipulations. A useful identity follows directly from its definition:
\begin{equation}
I_{\left(  K\right)  J}=\sum_{K^{\prime}\subseteq K}\left(  -1\right)
^{\left\vert K^{\prime}\right\vert +1}I_{K^{\prime}J},
\end{equation}
where $\left(  K\right)  $ denotes that the subset $K$ is treated as a single region. For example,
\begin{equation}
I_{12\left(  34\right)  } =S_{1}+S_{2}+S_{34}-S_{12}-S_{134}-S_{234}+S_{1234}
=I_{123}+I_{124}-I_{1234} \label{I-12(34)}%
\end{equation}
In general,
\begin{equation}
I_{\left(  K_{1}\right)  \cdots\left(  K_{m}\right)  }=\sum_{K_{i}^{\prime
}\subseteq K_{i}}\left(  -1\right)  ^{\sum_{i}\left(  \left\vert K_{i}%
^{\prime}\right\vert +1\right)  }I_{K_{1}^{\prime}\cdots K_{m}^{\prime}}.
\label{I-identity}%
\end{equation}


\subsection{HEIs in multipartite systems}

In a four-partite system, the independent non-redundant HEIs (up to the permutations of indices)
are
\begin{align}
I_{123}  &  \leq0,\\
I_{123}+I_{124}-I_{1234}  &  \leq0,
\end{align}
The first is the familiar MMI, and the second can be rewritten as
\begin{equation}
I_{123}+I_{124}-I_{1234} = I_{12\left(  34\right)  } \leq0,
\end{equation}
which is an enhanced version of MMI. Thus, there are no new inequalities in a four-partite system.

For a five-partite system, however, new independent HEIs appear. Five such non-redundant inequalities can be written as: 
\begin{align}
Q_{1}^{\left(  5\right)  }  &  =-I_{123}-I_{124}-I_{135}-I_{245}%
-I_{345}+I_{1234}+I_{1235}+I_{1245}+I_{1345}+I_{2345}-I_{12345},\\
Q_{2}^{\left(  5\right)  }  &  =-I_{123}-I_{124}-I_{125}-I_{345}%
+I_{1234}+I_{1235}+I_{1245},\\
Q_{3}^{\left(  5\right)  }  &  =-I_{123}-I_{124}-I_{135}-I_{245}%
+I_{1234}+I_{1235}+I_{1245},\\
Q_{4}^{\left(  5\right)  }  &  =-I_{123}-I_{124}-I_{125}-I_{345}%
+I_{1235}+I_{1245},\\
Q_{5}^{\left(  5\right)  }  &  =-I_{123}-2I_{125}-2I_{134}-I_{145}%
-I_{234}-I_{235}+2I_{1234}+2I_{1235}+I_{1245}+I_{1345}%
\end{align}
These inequalities cannot be expressed simply as combinations of MMI or its enhanced forms. The main goal of this work is to construct a systematic geometric method to prove them.

\subsection{Simplex Basis}

Each entangling region ${A}_{i}=[A_{i}^{L},A_{i}^{R}]$ is characterized by its
left and right endpoints on the boundary. We introduce the simplex basis ${S}_{\left[  ij\right]  }$, defined as the minimum surface connecting $A_{i}%
^{L}$ and $A_{j}^{R}$ in the bulk spacetime.

\begin{figure}[ptb]
\centering\includegraphics[width=.5\linewidth]{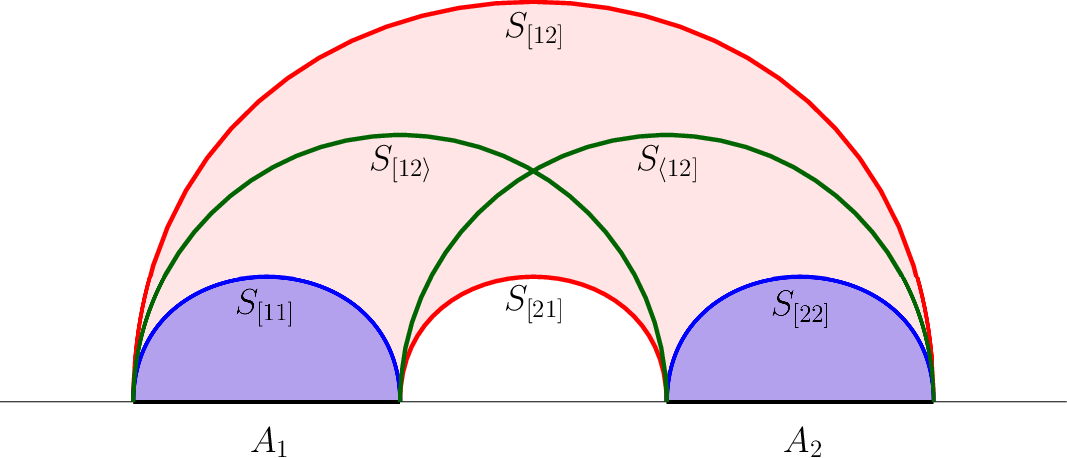}\caption{The definition of simplex basis. The six curves are the RT surfaces connecting the ends of entangling regions $A_1$ and $A_2$.}%
\label{FS12}%
\end{figure}

To illustrate, consider a simple bipartite system. Let ${A}_{ij}\equiv{A}_{i}\cup{A}_{j}$ be the union
of the two disjoint regions. The entanglement entropy $S_{12}$ can correspond to either a disconnected or a connected configuration of RT surfaces, as shown in Fig.\ref{FS12}. For the disconnected case, the total area is $S_{1,2}^{d}=S_{\left[  11\right]  }+S_{\left[  22\right]  }%
$, while for the connected case it is $S_{12}^{c}=S_{\left[
12\right]  }+S_{\left[  21\right]  }$. The physical entropy is the minimum of the two:
\begin{equation}
S_{12}=\min\left(  S_{1,2}^{d},S_{12}^{c}\right)  , \label{S12}%
\end{equation}
If $S_{1,2}^{d}\leq S_{12}^{c}$, the regions remain disconnected and $S_{12}=S_{1,2}^{d}=S_{1}+S_{2}$. If $S_{1,2}^{d}\geq S_{12}^{c}$, they form a connected configuration and $S_{12}=S_{12}^{c}\leq
S_{1,2}^{d}=S_{1}+S_{2}$. In either case, the subadditivity inequality eq.(\ref{SA}) is automatically satisfied, where equality happens when the two configurations have equal area.

In Fig.\ref{FS12}, we also define two auxiliary quantities, $S_{[12\rangle}$ and
$S_{\langle12]}$, which do not appear directly in the final inequalities but are useful as intermediate elements in the graphical proofs.

For a multipartite system, the connected-configuration entropy of a union region $A_{{k_{1}}...{k_{n}}}=\cup_{i=1}^{n}{A}_{k_{i}}$  can be decomposed as
\begin{equation}
S_{{k_{1}}...{k_{n}}}^{c}=\sum_{\sigma=1}^{n}S_{\left[  {k_{\sigma+1}%
k_{\sigma}}\right]  }, \label{expansion}%
\end{equation}
where ${k_{1}<k_{2}<}...<{k_{n}}$ and ${k_{n+1}\equiv k}_{1}$.

The set $\{{S}_{\left[  ij\right]  }\}$ forms the simplex basis, which will serve as the fundamental building blocks for the graphical analysis of holographic entanglement entropy inequalities in the following sections. 

\section{Tripartite System}

In this section, we consider a system composed of three disjoint entangled regions $\left(
A_{1},A_{2},A_{3}\right)$ satisfying $A_{i}\cap A_{j}=\emptyset$. Two fundamental holographic entropy inequalities apply to this system:  SSA and MMI.

\subsection{Strong Subadditivity (SSA)}

For the SSA in eq.(\ref{SSA}),
\begin{equation}
S_{12}+S_{23}\geq S_{123}+S_{2}, \label{SSAc}%
\end{equation}
we first examine a configuration where all terms correspond to connected entanglement wedges, which will be called the completely connected (CC) configuration. 

\begin{figure}[ptb]
\centering{ \subfloat[]{
\includegraphics[width=.4\linewidth]{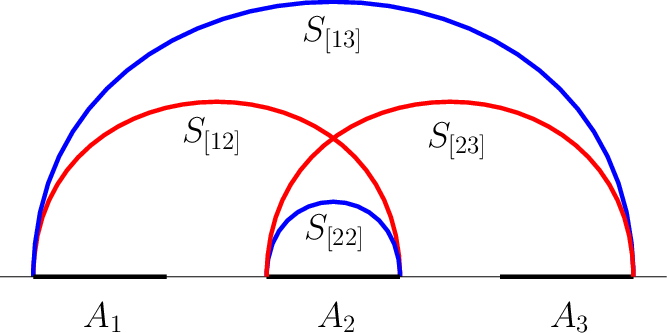} }\hspace{1cm}
\subfloat[]{
\includegraphics[width=.3\linewidth]{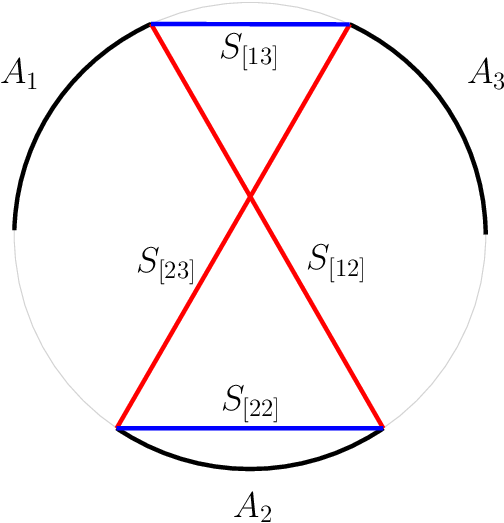} }} \caption{(a) The graphical proof of SSA. (b) Circular diagram of the SSA.}%
\label{fig-SSA}%
\end{figure}

In simplex basis, the individual entropies are written as
\begin{align}
S_{2}  &  =S_{\left[  22\right]  },\label{S12c}\\
S_{12}^{c}  &  =S_{\left[  12\right]  }+S_{\left[  21\right]  },\\
S_{23}^{c}  &  =S_{\left[  23\right]  }+S_{\left[  32\right]  },\label{S23c}\\
S_{123}^{c}  &  =S_{\left[  13\right]  }+S_{\left[  32\right]  }+S_{\left[
21\right]  }. \label{S123c}%
\end{align}
Substituting these expressions into eq.(\ref{SSAc}) gives the equivalent inequality
\begin{equation}
S_{\left[  12\right]  }+S_{\left[  23\right]  }\geq S_{\left[  13\right]
}+S_{\left[  22\right]  }. \label{SSAf}%
\end{equation}
To prove eq.(\ref{SSAf}), we compare the two red curves $S_{\left[
12\right]  }$ and $S_{\left[  23\right]  }$ with the blue curves $S_{\left[
13\right]  }$ and $S_{\left[  22\right]  }$ shown schematically in Fig.\ref{fig-SSA}(a). We cut the red curves $S_{\left[
12\right]  }$ and $S_{\left[  23\right]  }$ at their intersection point and reconnect them to form two new curves. One of these new curves is homologous to $S_{\left[  22\right]  }$, while the other is homologous to $S_{\left[  13\right]  }$. Since the RT surface minimizes the area within its homology class, each new curve must have an area greater than or equal to the corresponding minimal surface. Hence, the sum of the original red curves is necessarily greater than or equal to that of the blue curves, proving the inequality.

To simplify the visualization, we introduce a circular diagram in Fig.\ref{fig-SSA}(b), which encodes the connectivity structure of the RT surfaces. Each entanglement regions $A_{i}$ is placed along a circle in counterclockwise order, and each RT surface  $S_{\left[  ij\right]  }$ is represented as a straight line connecting $A_{i}$ and $A_{j}$. The inequality eq.(\ref{SSAf}) now corresponds to the statement that the sum of the two crossing red lines is greater than the sum of the two blue lines, which we call a cross inequality.

In general, a circular diagram representing an HEI satisfies two properties :

\begin{enumerate}

\item The numbers of red and blue lines are equal (ensuring balance). 

\item Each node $A_{i}$ connects to the same number of red and blue lines.
\end{enumerate}

These properties will be useful in later sections to prove more general inequalities.

Having proven SSA in the CC configuration, we now consider the case where some subsystems are disconnected. If $S_{12}$ is disconnected, i.e. $S_{12}=S_{1,2}%
^{d}=S_{1}+S_{2}$, the inequality reduces to
\begin{equation}
S_{1}+S_{23}\geq S_{123},
\end{equation}
which is simply the SA applied to the composite region ${A}_{23}$. The same reasoning applies if $S_{13}$ or $S_{23}$ is disconnected.

If $S_{123}$ is disconnected, four possible configurations arise:
\begin{align}
S_{1,23}^{d}  &  =S_{1}+S_{23}^{c}=S_{\left[  11\right]  }+S_{\left[
23\right]  }+S_{\left[  32\right]  },\label{1-23}\\
S_{12,3}^{d}  &  =S_{12}^{c}+S_{3}=S_{\left[  12\right]  }+S_{\left[
21\right]  }+S_{\left[  33\right]  },\label{12-3}\\
S_{2,31}^{d}  &  =S_{2}^{c}+S_{13}=S_{\left[  22\right]  }+S_{\left[
13\right]  }+S_{\left[  31\right]  },\label{2-31}\\
S_{1,2,3}^{d}  &  =S_{1}+S_{2}+S_{3}=S_{\left[  11\right]  }+S_{\left[
22\right]  }+S_{\left[  33\right]  }. \label{1-2-3}%
\end{align}
Let us examine these one by one. For the configuration eq.(\ref{1-23}), the definition of the minimal surface immediately gives
\begin{equation}
S_{123}^{c}\geq S_{1}+S_{23}^{c},
\end{equation}
which expands to
\begin{equation}
S_{\left[  13\right]  }+S_{\left[  32\right]  }+S_{\left[  21\right]  }\geq
S_{\left[  11\right]  }+S_{\left[  23\right]  }+S_{\left[  32\right]  },
\end{equation}
or equivalently,
\begin{equation}
S_{\left[  13\right]  }+S_{\left[  21\right]  }\geq S_{\left[  11\right]
}+S_{\left[  23\right]  }. \label{1-23f}%
\end{equation}
Adding this to eq.(\ref{SSAf}) yields
\begin{equation}
S_{\left[  13\right]  }+S_{\left[  21\right]  }+S_{\left[  12\right]
}+S_{\left[  23\right]  }\geq S_{\left[  11\right]  }+S_{\left[  23\right]
}+S_{\left[  13\right]  }+S_{\left[  22\right]  },
\end{equation}
which simplifies to
\begin{equation}
S_{12}^{c}=S_{\left[  12\right]  }+S_{\left[  21\right]  }\geq S_{\left[
11\right]  }+S_{\left[  22\right]  }=S_{1,2}^{d}.
\end{equation}
Hence, the assumption $S_{123}=S_{1,23}^{d}$ implies $S_{12}=S_{1,2}^{d}$, and SSA becomes a
trivial identity:
\begin{equation}
\left(  S_{1}+S_{2}\right)  +S_{23}^{c}=\left(  S_{1}+S_{23}^{c}\right)
+S_{2}.
\end{equation}
Similarly, $S_{123}=S_{12,3}^{d}$ implies $S_{23}=S_{2,3}^{d}$, and SSA becomes an identity
\begin{equation}
S_{12}^{c}+\left(  S_{2}+S_{3}\right)  =\left(  S_{12}^{c}+S_{3}\right)
+S_{2}.
\end{equation}
$S_{123}=S_{2,31}^{d}$ implies both $S_{12}=S_{1,2}^{d}$ and $S_{23}=S_{2,3}^{d}$, and the SSA becomes
\begin{equation}
\left(  S_{1}+S_{2}\right) +\left(  S_{2}+S_{3}\right)  \geq\left(  S_{13}%
^{c}+S_{2}\right)  +S_{2}.
\end{equation}

Finally, $S_{123}=S_{1,2,3}^{d}$ implies both $S_{12}=S_{1,2}^{d}$ and $S_{23}=S_{2,3}^{d}$, and the SSA implies a trivial identity
\begin{equation}
\left(  S_{1}+S_{2}\right)  +\left(  S_{2}+S_{3}\right)  =\left(  S_{1}%
+S_{2}+S_{3}\right)  +S_{2}.
\end{equation}
Thus, SSA eq.(\ref{SSA}) holds in all possible configurations.

\subsection{Monogamy of Mutual Information (MMI)}

\begin{figure}[ptb]
\centering{
\subfloat[]{\includegraphics[width=.25\linewidth]{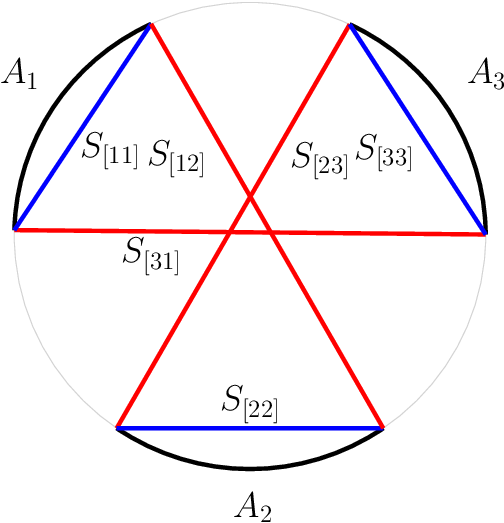} }\hspace
{.5cm}
\subfloat[]{\includegraphics[width=.25\linewidth]{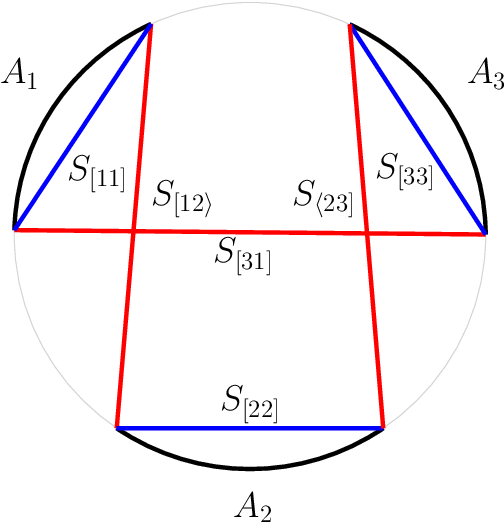} }\hspace
{.5cm} \subfloat[]{\includegraphics[width=.25\linewidth]{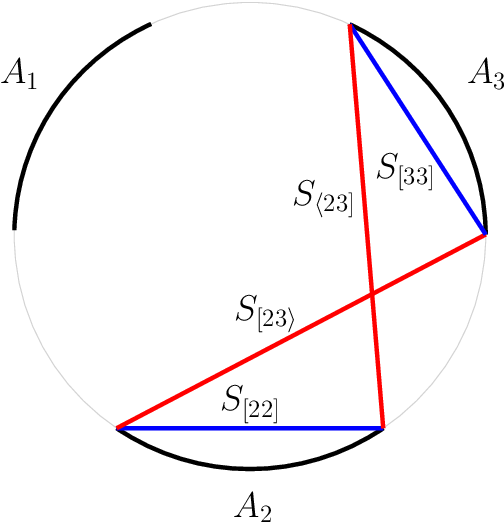} }}
\caption{Proof of the MMI through the "clean-gap procedure", shown step by step from (a) to (c) using cross inequalities.}%
\label{fig-MMI}%
\end{figure}

Next, we turn to the MMI eq.(\ref{MMI}),
\begin{equation}
S_{12}+S_{13}+S_{23}\geq S_{1}+S_{2}+S_{3}+S_{123}.
\label{MMIc}%
\end{equation}
We again begin with the completely connected configuration, in which the inequality can be rewritten in the simplex basis as 
\begin{equation}
S_{\left[  12\right]  }+S_{\left[  23\right]  }+S_{\left[  31\right]  }\geq
S_{\left[  11\right]  }+S_{\left[  22\right]  }+S_{\left[  33\right]  }.
\label{MMIfa}%
\end{equation}
To prove eq.(\ref{MMIfa}), consider Fig.\ref{fig-MMI}(a). The
three red lines correspond to the connected RT surfaces $S_{[12]}$, $S_{[23]}$ and $S_{[31]}$; the blue lines represent $S_{[11]}$, $S_{[22]}$ and $S_{[33]}$. Applying the cross inequality to $S_{[12]}$ and $S_{[23]}$,
\begin{equation}
S_{\left[  12\right]  }+S_{\left[  23\right]  }\geq S_{\left[  12\right\rangle
}+S_{\left\langle 23\right]  },
\end{equation}
reduces eq.(\ref{MMIfa}) to
\begin{equation}
S_{\left[  12\right\rangle }+S_{\left\langle 23\right]  }+S_{\left[
31\right]  }\geq S_{\left[  11\right]  }+S_{\left[  22\right]  }+S_{\left[
33\right]  }.
\end{equation}
Now we need to show that the sum of the three red lines is greater than the
sum of the three blue lines in Fig.\ref{fig-MMI}(b).

Next, applying the cross inequality to $S_{[12\rangle}$ and $S_{[31]}$,
\begin{equation}
S_{\left[  12\right\rangle }+S_{\left[  31\right]  }\geq S_{\left[  11\right]
}+S_{\left[  23\right\rangle }, \label{MMIfb}%
\end{equation}
yields
\begin{equation}
S_{\left[  23\right\rangle }+S_{\left\langle 23\right]  }\geq S_{\left[
22\right]  }+S_{\left[  33\right]  }. \label{MMIfc}%
\end{equation}
The final relation, illustrated in Fig.\ref{fig-MMI}(c), is another example of the cross inequality between two red and two blue lines. Therefore, the inequality eq.(\ref{MMIc}) is proven in the completely connected configuration.

We call this sequence of operations, i.e., successively eliminating crossed lines by exchanging their endpoints across each gap, the "clean-gap" procedure. It is a key step in all graphical proofs of HEIs.

Now consider other configurations. If one of the $S_{ij}$ is
disconnected, $S_{ij}=S_{i,j}^{d}=S_{i}+S_{j}$, MMI reduces to SSA. If more than one of them is disconnected, MMI reduce further to SA. On the other hand, if $S_{123}$ is disconnected, the four possible forms in eqs.(\ref{1-23}-\ref{1-2-3}) again apply.

It can be verified that
\begin{align}
S_{123}  &  =S_{1,23}^{d}\Rightarrow S_{12}=S_{1,2}^{d}\text{ and }%
S_{13}=S_{1,3}^{d},\\
S_{123}  &  =S_{12,3}^{d}\Rightarrow S_{13}=S_{1,3}^{d}\text{ and }%
S_{23}=S_{2,3}^{d},\\
S_{123}  &  =S_{2,31}^{d}\Rightarrow S_{12}=S_{1,2}^{d}\text{ and }%
S_{23}=S_{2,3}^{d},\\
S_{123}  &  =S_{1,2,3}^{d}\Rightarrow S_{12}=S_{1,2}^{d},S_{13}=S_{1,3}%
^{d},\text{ and }S_{23}=S_{2,3}^{d}.
\end{align}
In each of the these cases, MMI reduces to a trivial identity, confirming that MMI holds universally.

The examples above illustrate a general strategy for proving holographic entropy inequalities: 

\begin{enumerate}
\item Express each inequality in the simplex basis under the completely connected configuration.

\item Use successive cross inequalities to prove it graphically.

\item Extend the proof to all other configurations by “cutting” connected terms.
\end{enumerate}

In the third step, we have shown that cutting one surface may require cutting others as well.\footnote{This is the most important difference between the $\left(1+1\right)$-dimensional CFT and higher dimensional CFT studied in
\cite{1808.07871,1812.08133,1905.06985,2002.04558}}. This observation motivates the Cut Theorem, which will be proved later as a general principle for extending proofs from connected to arbitrary configurations.

\section{Miltipartite System}

We now extend our graphical method to general \textit{n}-partite systems. The approach follows the same logic used for the tripartite case but introduces several new features. In particular, for $n\geq4$, certain configurations of connected RT surfaces become mutually inconsistent, leading us to define a compatibility condition among entanglement regions.

This section introduces the Compatible Theorem and the concept of Compatible Completely Connected (CCC) configurations, which form the foundation for proving superbalanced holographic entanglement entropy inequalities (HEIs) in higher-partite systems. 

\subsection{Compatible Theorem}

When $n\geq4$, not all configurations of RT surfaces are physically allowed. To illustrate, consider a $4$-partite system. Suppose that both $S_{13}=S_{13}^{c}$ and $S_{24}=S_{24}^{c}$ are
connected simultaneously. This would imply that
\begin{align}
S_{13}^{c}  &  =S_{\left[  13\right]  }+S_{\left[  31\right]  }\leq
S_{1,3}^{d}=S_{\left[  11\right]  }+S_{\left[  33\right]  },\\
S_{24}^{c}  &  =S_{\left[  24\right]  }+S_{\left[  42\right]  }\leq
S_{2,4}^{d}=S_{\left[  22\right]  }+S_{\left[  44\right]  }.
\end{align}
Adding the two inequalities gives the following result
\begin{equation}
S_{\left[  13\right]  }+S_{\left[  31\right]  }+S_{\left[  24\right]
}+S_{\left[  42\right]  }\leq S_{\left[  11\right]  }+S_{\left[  22\right]
}+S_{\left[  33\right]  }+S_{\left[  44\right]  }, \label{13-24}%
\end{equation}
i.e. the sum of the red lines (the connected surfaces) is less than the sum of the blue lines (the disconnected surfaces).

However, as we will show, the geometric constraints of the RT prescription imply the opposite inequality:
\begin{equation}
S_{\left[  13\right]  }+S_{\left[  31\right]  }+S_{\left[  24\right]
}+S_{\left[  42\right]  }\geq S_{\left[  11\right]  }+S_{\left[  22\right]
}+S_{\left[  33\right]  }+S_{\left[  44\right]  }. \label{13-24 oppo}%
\end{equation}
This contradiction means that the two surfaces $S_{13}^{c}$ and $S_{24}^{c}$ cannot be both connected at the same time; they are incompatible.

\begin{figure}[ptb]
\centering{
\subfloat[]{\includegraphics[width=.23\linewidth]{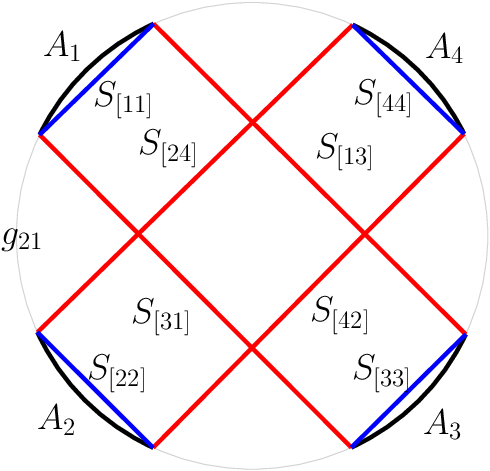} }\hspace
{.12cm}
\subfloat[]{\includegraphics[width=.23\linewidth]{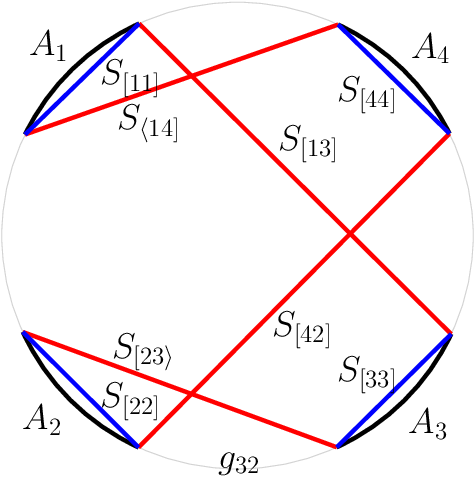} }\hspace
{.12cm}
\subfloat[]{\includegraphics[width=.23\linewidth]{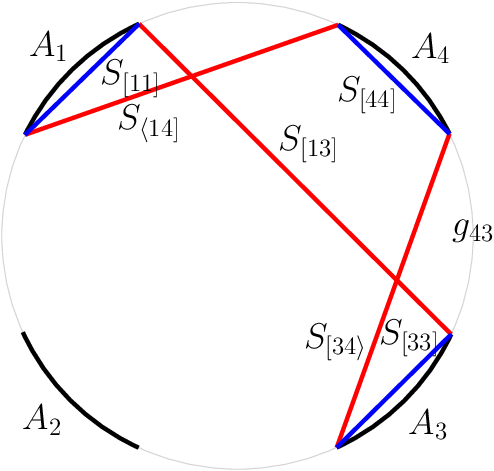} }\hspace
{.12cm}
\subfloat[]{\includegraphics[width=.23\linewidth]{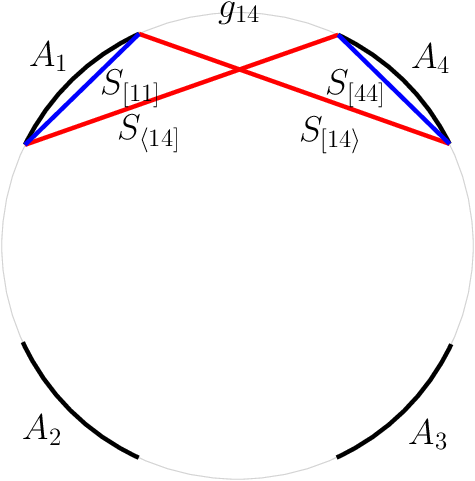} }\hspace
{.12cm}} \caption{Proving eq.(\ref{13-24}) by clean-gap procedure.}%
\label{fig-13-24}%
\end{figure}

To prove eq.(\ref{13-24 oppo}), we employ successive application of the cross inequality, which corresponds to exchanging the endpoints of intersecting red lines in the circular diagram 
Fig.\ref{fig-13-24}.

Then, using the cross inequality,%
\begin{equation}
S_{\left[  13\right]  }+S_{\left[  34\right\rangle }\geq S_{\left[
14\right\rangle }+S_{\left[  33\right]  }, \label{cross1334}%
\end{equation}
the inequality is reduced to%
\begin{equation}
S_{\left[  14\right\rangle }+S_{\left\langle 14\right]  }\geq S_{\left[
11\right]  }+S_{\left[  44\right]  }, \label{13-24 final}%
\end{equation}
where $S_{\left[  33\right]  }$ cancels out on both sides of the inequality.
The circular diagram is shown in Fig.\ref{fig-13-24}(d).

Finally, we notice that the inequality eq.(\ref{13-24 final}) itself is a
cross inequalities. Thus, we have verified the inequality eq.(\ref{13-24 oppo})
which is in contradiction to eq.(\ref{13-24}). We therefore conclude that
$S_{13}$ and $S_{24}$ cannot be connected simultaneously. We say that
$S_{13}^{c}$ and $S_{24}^{c}$ are incompatible.

Let's review how we proved that $S_{13}^{c}$ and $S_{24}^{c}$ are incompatible
by looking at their circular diagrams step by step:

\begin{itemize}
\item First, in Fig.\ref{fig-13-24}(a), we observe that the red lines $S_{\left[  31\right]
}$ and $S_{\left[  24\right]  }$ cross each other. Exchanging their endpoints gives two new red lines $S_{\left[  23\right\rangle
}$ and $S_{\left\langle 14\right]  }$, and the inequality eq.(\ref{13-24 oppo}) reduces to
\begin{equation}
S_{\left[  13\right]  }+S_{\left[  23\right\rangle }+S_{\left\langle
14\right]  }+S_{\left[  42\right]  }\geq S_{\left[  11\right]  }+S_{\left[
22\right]  }+S_{\left[  33\right]  }+S_{\left[  44\right]  }.
\end{equation}
This operation (called "skipping the gap $g_{21}$") produces a new configuration Fig.\ref{fig-13-24}(b) that is smaller according to the cross inequality
\begin{equation}
S_{\left[  31\right]  }+S_{\left[  24\right]  }\geq S_{\left[  23\right\rangle
}+S_{\left\langle 14\right]  }. \label{cross3124}%
\end{equation}

\item Next, in Fig.\ref{fig-13-24}(b), the red lines $S_{\left[  23\right\rangle }$ and $S_{\left[  42\right]
}$ cross each other. Exchanging their endpoints by skipping the gap $g_{32}$ yields $S_{\left[  22\right]  }$ and $S_{\left[  34\right\rangle }$, and the inequality further reduces to
\begin{equation}
S_{\left[  13\right]  }+S_{\left\langle 14\right]  }+S_{\left[
34\right\rangle }\geq S_{\left[  11\right]  }+S_{\left[  33\right]
}+S_{\left[  44\right]  },
\end{equation}
where term $S_{\left[  22\right]  }$ cancels on both sides. The new configuration Fig.\ref{fig-13-24}(c) is smaller according to the cross inequality
\begin{equation}
S_{\left[  23\right\rangle }+S_{\left[  42\right]  }\geq S_{\left[  22\right]
}+S_{\left[  34\right\rangle }. \label{cross2342}%
\end{equation}

\item Then, in Fig.\ref{fig-13-24}(c), the red lines $S_{\left[  13\right]  }$ and $S_{\left[  34\right\rangle
}$ cross each other. Exchanging their endpoints by skipping the gap $g_{43}$ yields $S_{\left[  14\right\rangle }$ and $S_{\left[  33\right]  }$, and the inequality reduces to
\begin{equation}
S_{\left[  14\right\rangle }+S_{\left\langle 14\right]  }\geq S_{\left[
11\right]  }+S_{\left[  44\right]  }, \label{13-24 final}
\end{equation}

where term $S_{\left[  33\right]  }$ cancels on both sides. The new configuration Fig.\ref{fig-13-24}(d) is smaller according to the cross inequality\begin{equation}
S_{\left[  13\right]  }+S_{\left[  34\right\rangle }\geq S_{\left[
14\right\rangle }+S_{\left[  33\right]  }. \label{cross1334}%
\end{equation}

\item Finally, in Fig.\ref{fig-13-24}(d), the pair $S_{\left[  14\right\rangle }$ and $S_{\left\langle 14\right]  }$ cross each other. Exchanging their endpoints by skipping the gap $g_{43}$ yields $S_{\left[  11\right]  }$ and $S_{\left[  44\right]  }$, both of which cancel the
original blue lines. This chain of exchanges completes the proof of inequality eq.(\ref{13-24 oppo}), confirming that $S_{13}^{c}$ and $S_{24}^{c}$ cannot coexist.
\end{itemize}

Generally, we have the following Theorem.

\noindent\textbf{Compatible Theorem}: In a multipartite system, two connected RT surfaces $S_{X}$ and $S_{X'}$ are incompatible if the following conditions hold:

\begin{enumerate}
\item Complementarity: $X\cap X'=\emptyset,$ i.e. the two subsystems do not share boundary regions. 

\item Interlaced: the RT surfaces of $S_{X}$ and $S_{X^{\prime}}$ are non-planar and cross each other in the circular diagram.
\end{enumerate}

\noindent\textbf{Proof}: Arrange $X$ and $X^{\prime}$ in alternating ordered blocks along the circle,
\begin{equation}
X_{1}X^{\prime}{}_{1}X_{2}X^{\prime}{}_{2}\cdots X_{m}X^{\prime}{}_{m}%
\end{equation}
where $X=X_{1}X_{2}\cdots X_{m}$, $X^{\prime}=X_{1}^{\prime}X^{\prime}{}%
_{2}\cdots X_{m}^{\prime}$ and $m$ counts the number of alternations. 

\begin{figure}[ptb]
\centering{ \includegraphics[width=.35\linewidth]{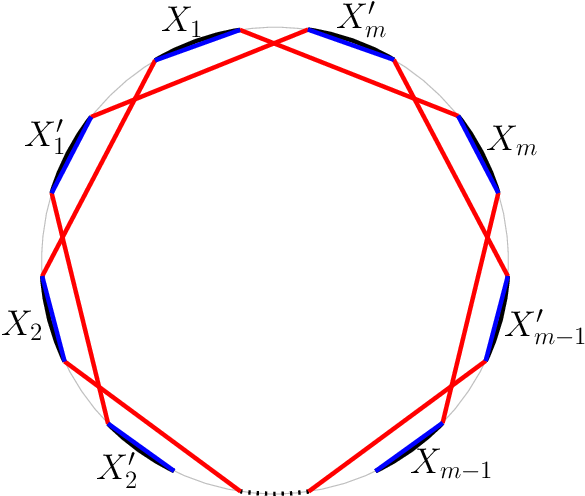}
}\caption{Proof of Compatible Theorem: The sum of the red lines is larger than the sum of the blue lines by the clean-gap procedure, which is in contradiction to eq.(\ref{ip-qj}).}%
\label{incompatible}%
\end{figure}

Assuming both $S_{X}^{c}$ and $S_{X^{\prime}}^{c}$ are connected, their areas satisfy
\begin{align}
S_{X}^{c}  &  =S_{X_{1}\cdots X_{m}}^{c}\leq\sum_{k=1}^{m}S_{X_{k}}%
^{c},\label{SXc}\\
S_{X^{\prime}}^{c}  &  =S_{X_{1}^{\prime}\cdots X_{m}^{\prime}}^{c}\leq
\sum_{k=1}^{m}S_{X_{k}^{\prime}}^{c}. \label{SX'c}%
\end{align}
Using eq.(\ref{expansion}), each term expands in the simplex basis as
\begin{align}
S_{X}^{c}  &  =\sum_{k=1}^{m}\left(  S_{\left[  \left( k+1\right) _{1}%
k_{n_{k}}\right]  }+\sum_{\sigma=1}^{n_{k}-1}S_{\left[  {k_{\sigma+1}%
k_{\sigma}}\right]  }\right)  ,\sum_{k=1}^{m}S_{X_{k}}^{c}=\sum_{k=1}^{m}%
\sum_{\sigma=1}^{n_{k}}S_{\left[  {k_{\sigma+1}k_{\sigma}}\right]  },\\
S_{X^{\prime}}^{c}  &  =\sum_{k=1}^{m}\left(  S_{\left[  \left( k^{\prime
}+1\right) _{1}k^{\prime}_{n_{k^{\prime}}}\right]  }\sum_{\sigma=1}%
^{n_{k}^{\prime}-1}S_{\left[  {k}_{\sigma+1}^{\prime}{k}_{\sigma}^{\prime
}\right]  }\right)  ,\sum_{k=1}^{m}S_{X_{k}^{\prime}}^{c}=\sum_{k^{\prime}%
=1}^{m}\sum_{\sigma=1}^{n_{k}^{\prime}}S_{\left[  {k}_{\sigma+1}^{\prime}%
{k}_{\sigma}^{\prime}\right]  }.
\end{align}
After canceling common terms, the inequalities eq.(\ref{SXc}-\ref{SX'c}) reduce to
\begin{align}
\sum_{k=1}^{m}S_{\left[  \left( k+1\right) _{1} k_{n_{k}}\right]  }  &
\leq\sum_{k=1}^{m}S_{\left[  {k_{1}k_{n_{k}}}\right]  }\\
\sum_{k=1}^{m}S_{\left[  \left( k^{\prime}+1\right) _{1} k^{\prime
}_{n_{k^{\prime}}}\right]  }  &  \leq\sum_{k=1}^{m}S_{\left[  {k}_{1}^{\prime
}{k}_{n_{k^{\prime}}}^{\prime}\right]  }.
\end{align}
Adding the two inequalities gives
\begin{equation}
\sum_{k=1}^{m}\left(  S_{\left[  \left( k+1\right) _{1}k_{n_{k}}\right]
}+S_{\left[  \left( k+1\right) _{1}^{\prime} k^{\prime}_{n_{k^{\prime}}%
}\right]  }\right)  \leq\sum_{k=1}^{m}\left(  S_{\left[  {k_{1}k_{n_{k}}%
}\right]  }+S_{\left[  {k}_{1}^{\prime}{k}_{n_{k^{\prime}}}^{\prime}\right]
}\right)  ,
\label{ip-qj}
\end{equation}
which means that the sum of the red lines is smaller than the sum of the blue
lines in Fig.\ref{incompatible}.

However, applying the clean-gap procedure reverses the inequality eq.(\ref{ip-qj}), leading to a contradiction.

Hence, $S_{X}$ and $S_{X^{\prime}}$ cannot be connected simultaneously, i.e., $S_{X}%
^{c}$ and $S_{X^{\prime}}^{c}$ are incompatible. $\Box$

\subsection{Compatible Completely Connected Configuration}

The Compatible Theorem imposes strong restrictions for the allowed configurations of a HEI.

In the tripartite case, all terms could be connected simultaneously. However, in a 4-partite system, $S_{13}^{c}$ and $S_{24}^{c}$ are incompatible, so at least one of them must be disconnected.

Configurations that satisfy all such compatibility constraints are called Compatible Completely Connected (CCC) configurations. In a 4-partite system, there are two CCC configurations: one with $S_{13}$ disconnected and one with $S_{24}$ disconnected.

In a 5-partite system, compatibility becomes more restrictive. There are five pairs of incompatible RT surfaces as shown in Table \ref{CCC5}. At least one surface in each pair must be disconnected.
\begin{table}[ptb]
\begin{center}%
\begin{tabular}
[c]{|l|l|l|l|l|l|}\hline
& $S_{124}$ & $S_{134}$ & $S_{135}$ & $S_{235}$ & $S_{245}$\\\hline
$S_{13}$ &  &  &  &  & $\checkmark$\\\hline
$S_{14}$ &  &  &  & $\checkmark$ & \\\hline
$S_{24}$ &  &  & $\checkmark$ &  & \\\hline
$S_{25}$ &  & $\checkmark$ &  &  & \\\hline
$S_{35}$ & $\checkmark$ &  &  &  & \\\hline
\end{tabular}
\end{center}
\caption{Incompatible pairs are checked in the 5-partite system. One of the RT surfaces in the incompatible pairs must be disconnected.}%
\label{CCC5}%
\end{table}

Accounting for all combinations, there are eleven allowed CCC configurations. Each includes exactly five disconnected terms, either explicitly chosen or implied by others through the constraints demonstrated earlier. We list all allowed CCC configurations in the following:
\begin{align}
\text{CCC}1 &  :S_{1,3}^{d},S_{1,4}^{d},S_{2,4}^{d},S_{2,5}^{d},S_{3,5}^{d}\\
\text{CCC}2 &  :S_{1,3}^{d},S_{2,5}^{d},S_{12,4}^{d}\Rightarrow S_{1,4}%
^{d},S_{2,4}^{d}\\
\text{CCC}3 &  :S_{2,4}^{d},S_{3,5}^{d},S_{1,34}^{d}\Rightarrow S_{1,3}%
^{d},S_{1,4}^{d}\\
\text{CCC}4 &  :S_{1,4}^{d},S_{2,5}^{d},S_{3,51}^{d}\Rightarrow S_{1,3}%
^{d},S_{3,5}^{d}\\
\text{CCC}5 &  :S_{1,3}^{d},S_{2,4}^{d},S_{23,5}^{d}\Rightarrow S_{2,5}%
^{d},S_{3,5}^{d}\\
\text{CCC}6 &  :S_{1,4}^{d},S_{3,5}^{d},S_{2,45}^{d}\Rightarrow S_{2,4}%
^{d},S_{2,5}^{d}\\
\text{CCC}7 &  :S_{12,4}^{d},S_{1,34}^{d}\Rightarrow S_{1,3}^{d},S_{1,4}%
^{d},S_{2,4}^{d}\\
\text{CCC}8 &  :S_{1,34}^{d},S_{3,51}^{d}\Rightarrow S_{1,3}^{d},S_{1,4}%
^{d},S_{3,5}^{d}\\
\text{CCC}9 &  :S_{3,51}^{d},S_{23,5}^{d}\Rightarrow S_{1,3}^{d},S_{3,5}%
^{d},S_{2,5}^{d}\\
\text{CCC}10 &  :S_{12,4}^{d},S_{2,45}^{d}\Rightarrow S_{2,4}^{d},S_{2,5}%
^{d},S_{1,4}^{d}\\
\text{CCC}11 &  :S_{23,5}^{d},S_{2,45}^{d}\Rightarrow S_{2,5}^{d},S_{3,5}%
^{d},S_{2,4}^{d}%
\end{align}
where a certain disconnected term implies that other terms are disconnected. For example, in CCC2, $S_{12,4}^{d}$ implies $S_{1,4}^{d}$ and $S_{2,4}^{d}$ are disconnected.

\subsection{Proof of Superbalanced HEIs in CCC Configurations}

To prove a HEI in a $n$-partite system, one must verify it for all possible configurations. However, from our earlier examples, SSA and MMI, we have seen that once an inequality
holds in all CCC configuration, it automatically holds in every other configuration. We therefore focus on CCC configurations first.

\begin{figure}[ptb]
\centering{ \includegraphics[width=.35\linewidth]{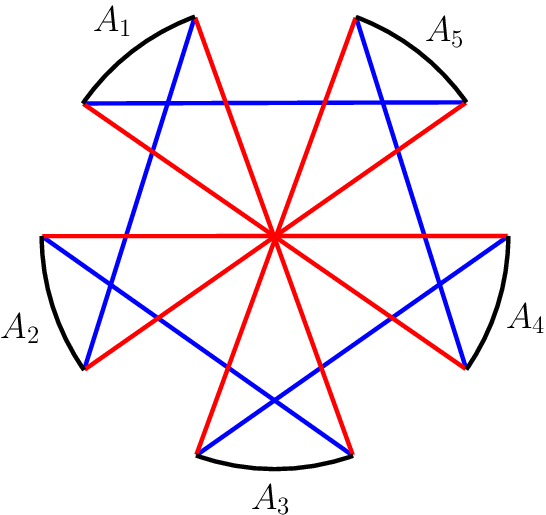}
}\caption{The circular diagram for of $Q_{1}^{\left(  5\right)  }$ in any CCC configuration.}%
\label{fig-Q1}%
\end{figure}

The first example we consider is a 5-partite HEI in Table 1 of \cite{2002.04558}, 
\begin{equation}
Q_{1}^{\left(  5\right)  }=-I_{124}-I_{134}-I_{135}-I_{235}-I_{245}%
+I_{1234}+I_{1235}+I_{1245}+I_{1345}+I_{2345}-I_{12345}\geq0.
\end{equation}
In the S-basis, this becomes
\begin{equation}
S_{123}+S_{234}+S_{345}+S_{145}+S_{125}\geq S_{12}+S_{23}+S_{34}+S_{45}%
+S_{15}+S_{12345}. \label{Q1}%
\end{equation}
Because all terms here are mutually compatible, we can choose every term to be connected in any CCC configuration:
\begin{equation}
S_{123}^{c}+S_{234}^{c}+S_{345}^{c}+S_{145}^{c}+S_{125}^{c}\geq S_{12}%
^{c}+S_{23}^{c}+S_{34}^{c}+S_{45}^{c}+S_{15}^{c}+S_{12345}^{c}. \label{Q1c}%
\end{equation}
Expressing this in the simplex basis gives
\begin{equation}
S_{\left[  13\right]  }+S_{\left[  24\right]  }+S_{\left[  35\right]
}+S_{\left[  41\right]  }+S_{\left[  52\right]  }\geq S_{\left[  12\right]
}+S_{\left[  23\right]  }+S_{\left[  34\right]  }+S_{\left[  45\right]
}+S_{\left[  51\right]  }.
\end{equation}
The corresponding circular diagram Fig.\ref{fig-Q1} can be simplified via the clean-gap
procedure, showing that the left-hand side is always greater. Thus, $Q_{1}^{\left(  5\right)  }$ holds in all CCC
configurations.

The next example is
\begin{equation}
Q_{2}^{\left(  5\right)  }=-I_{124}-I_{125}-I_{135}-I_{234}+I_{1234}%
+I_{1235}+I_{1245}\geq0,
\end{equation}
which in S-basis reads
\begin{equation}
2S_{123}+S_{124}+S_{125}+S_{134}+S_{145}+S_{235}+S_{245}\geq S_{12}%
+S_{13}+S_{14}+S_{23}+S_{25}+S_{45}+S_{1234}+S_{1235}+S_{1245}.\label{Q52}%
\end{equation}
We can now test this inequality under each CCC configuration by expressing all terms in the simplex basis.

\begin{enumerate}
\item In the configuration CCC$1$, eq.(\ref{Q52}) becomes%
\begin{equation}
2S_{\left[  13\right]  }+S_{\left[  14\right]  }+2S_{\left[  25\right]
}+S_{\left[  31\right]  }+S_{\left[  41\right]  }+S_{\left[  42\right]
}+S_{\left[  52\right]  }\geq2S_{\left[  11\right]  }+S_{\left[  12\right]
}+S_{\left[  22\right]  }+S_{\left[  23\right]  }+S_{\left[  33\right]
}+S_{\left[  44\right]  }+S_{\left[  45\right]  }+S_{\left[  55\right]  }.
\end{equation}

\item In the configuration CCC$2$, eq.(\ref{Q52}) becomes%
\begin{equation}
2S_{\left[  13\right]  }+2S_{\left[  25\right]  }+S_{\left[  31\right]
}+S_{\left[  41\right]  }+S_{\left[  52\right]  }\geq2S_{\left[  11\right]
}+S_{\left[  22\right]  }+S_{\left[  23\right]  }+S_{\left[  33\right]
}+S_{\left[  45\right]  }+S_{\left[  55\right]  }.
\end{equation}

\item In the configuration CCC$3$, eq.(\ref{Q52}) becomes%
\begin{equation}
2S_{\left[  13\right]  }+S_{\left[  25\right]  }+S_{\left[  34\right]
}+S_{\left[  41\right]  }+S_{\left[  42\right]  }\geq S_{\left[  11\right]
}+S_{\left[  12\right]  }+S_{\left[  23\right]  }+S_{\left[  33\right]
}+S_{\left[  44\right]  }+S_{\left[  45\right]  }.
\end{equation}

\item In the configuration CCC$4$, eq.(\ref{Q52}) becomes%
\begin{equation}
2S_{\left[  13\right]  }+S_{\left[  14\right]  }+2S_{\left[  25\right]
}+S_{\left[  31\right]  }+S_{\left[  41\right]  }+S_{\left[  42\right]
}+S_{\left[  52\right]  }\geq2S_{\left[  11\right]  }+S_{\left[  12\right]
}+S_{\left[  22\right]  }+S_{\left[  23\right]  }+S_{\left[  33\right]
}+S_{\left[  44\right]  }+S_{\left[  45\right]  }+S_{\left[  55\right]  },
\end{equation}
which is the same as the first CCC configuration.

\item In the configuration CCC$5$, eq.(\ref{Q52}) becomes%
\begin{equation}
2S_{\left[  13\right]  }+S_{\left[  25\right]  }+S_{\left[  31\right]
}+S_{\left[  42\right]  }+S_{\left[  52\right]  }\geq S_{\left[  11\right]
}+S_{\left[  12\right]  }+S_{\left[  22\right]  }+S_{\left[  33\right]
}+S_{\left[  45\right]  }+S_{\left[  53\right]  },
\end{equation}

\item In the configuration CCC$6$, eq.(\ref{Q52}) becomes%
\begin{equation}
S_{\left[  13\right]  }+S_{\left[  14\right]  }+S_{\left[  25\right]
}+S_{\left[  41\right]  }+S_{\left[  52\right]  }\geq S_{\left[  11\right]
}+S_{\left[  12\right]  }+S_{\left[  23\right]  }+S_{\left[  44\right]
}+S_{\left[  55\right]  }.
\end{equation}

\item In the configuration CCC$7$, eq.(\ref{Q52}) becomes%
\begin{equation}
2S_{\left[  13\right]  }+S_{\left[  25\right]  }+S_{\left[  34\right]
}+S_{\left[  41\right]  }\geq S_{\left[  11\right]  }+S_{\left[  14\right]
}+S_{\left[  23\right]  }+S_{\left[  33\right]  }+S_{\left[  45\right]  }.
\end{equation}

\item In the configuration CCC$8$, eq.(\ref{Q52}) becomes%
\begin{equation}
2S_{\left[  13\right]  }+S_{\left[  25\right]  }+S_{\left[  34\right]
}+S_{\left[  41\right]  }+S_{\left[  42\right]  }\geq S_{\left[  11\right]
}+S_{\left[  12\right]  }+S_{\left[  23\right]  }+S_{\left[  33\right]
}+S_{\left[  44\right]  }+S_{\left[  45\right]  },
\end{equation}
which is the same as the thrid CCC configuration.

\item In the configuration CCC$9$, eq.(\ref{Q52}) becomes%
\begin{equation}
2S_{\left[  13\right]  }+S_{\left[  25\right]  }+S_{\left[  31\right]
}+S_{\left[  42\right]  }+S_{\left[  52\right]  }\geq S_{\left[  11\right]
}+S_{\left[  12\right]  }+S_{\left[  22\right]  }+S_{\left[  33\right]
}+S_{\left[  45\right]  }+S_{\left[  53\right]  },
\end{equation}
which is the same as the fifth CCC configuration.

\item In the configuration CCC$10$, eq.(\ref{Q52}) becomes%
\begin{equation}
S_{\left[  13\right]  }+S_{\left[  25\right]  }+S_{\left[  41\right]
}+S_{\left[  52\right]  }\geq S_{\left[  11\right]  }+S_{\left[  23\right]
}+S_{\left[  42\right]  }+S_{\left[  55\right]  }.
\end{equation}

\item In the configuration CCC$11$, eq.(\ref{Q52}) becomes%
\begin{equation}
S_{\left[  13\right]  }+S_{\left[  52\right]  }\geq S_{\left[  12\right]
}+S_{\left[  53\right]  }.
\end{equation}

\end{enumerate}

Although there are eleven CCC configurations, some of them are equivalent, e.g. CCC1 and CCC4 yield the same reduced form.

\begin{figure}[ptb]
\centering{
\subfloat[CCC1 and CCC4]{\includegraphics[width=.3\linewidth]{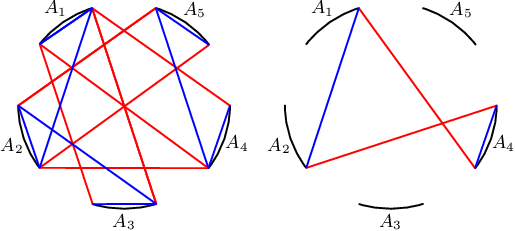} }\hspace
{.2cm}
\subfloat[CCC2]{\includegraphics[width=.3\linewidth]{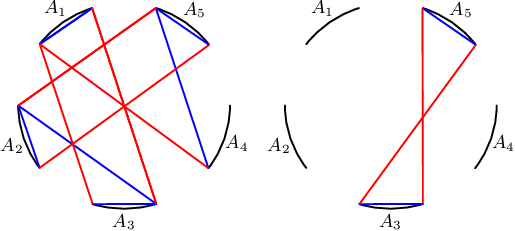} }\hspace
{.2cm}
\subfloat[CCC3 and CCC8]{\includegraphics[width=.3\linewidth]{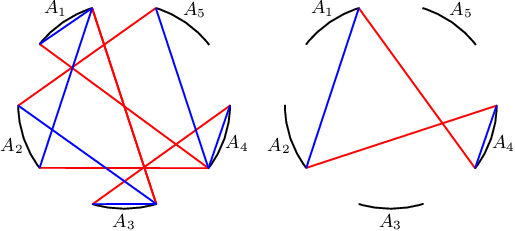} }\newline%
\subfloat[CCC5 and CCC9]{\includegraphics[width=.3\linewidth]{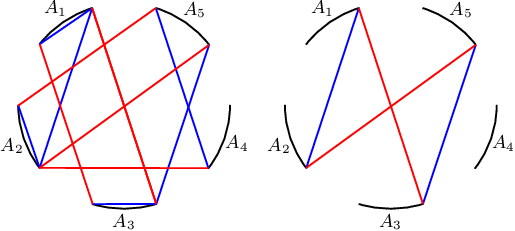}}\hspace
{.2cm}
\subfloat[CCC6]{\includegraphics[width=.3\linewidth]{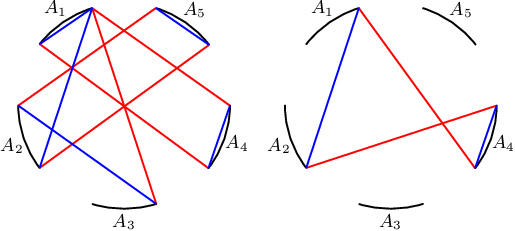}}\hspace
{.2cm}
\subfloat[CCC7]{\includegraphics[width=.3\linewidth]{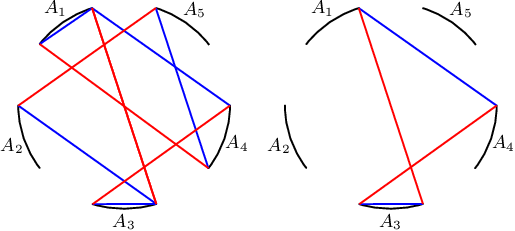}}\newline%
\subfloat[CCC10]{\includegraphics[width=.3\linewidth]{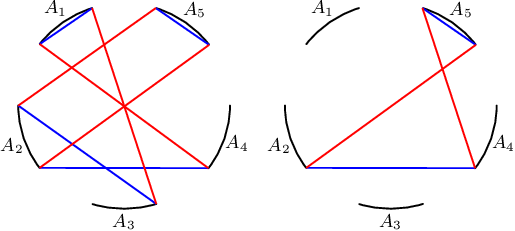}}\hspace
{.2cm}
\subfloat[CCC11]{\includegraphics[width=.3\linewidth]{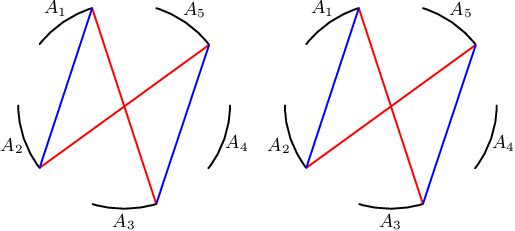}}\hspace
{.2cm}
\subfloat[Joint form of $Q_{2}^{5}$]{\includegraphics[width=.3\linewidth]{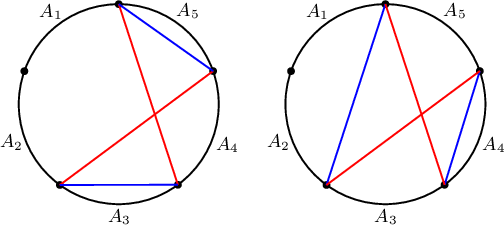}}\hspace
{.2cm} } \caption{The graphical proof of $Q_{2}^{5}$ in 11 CCC configurations. The joint circular diagram of $Q_{2}^{5}$ is shown in (i).}%
\label{fig-Q52}%
\end{figure}

In each case, the resulting circular diagram (shown in Fig.\ref{fig-Q52}) reduces to a positive combination of cross inequalities, confirming that $Q_{2}^{\left(  5\right)  }\ge 0$ in every CCC configuration.

Let's summarize the strategy we used to prove a HEI in a CCC configuration:

\begin{enumerate}
\item Express the HEI in the simplex basis and plot the circular diagrams.

\item Reduce the circular diagrams by the clean-gap procedure.

\item Prove the reduced diagrams by cross inequalities.
\end{enumerate}

\subsection{Joint Form}

The strategy developed in the previous section provides a direct way to prove HEIs in CCC configurations. However, verifying every CCC configuration one by one is often repetitive and inefficient. 

To simplify the procedure, we note that the reduced circular diagrams obtained after applying the\textit{ }clean-gap procedure exhibit a common internal structure: the pattern of red and blue lines in each reduced circular diagram is largely independent of the specific region labels  $A_{i}$.
More precisely, if a circular diagram contains no RT surfaces that connect the gaps, i.e. gapless, these gaps can be closed without changing the structure of the inequality.  When the adjacent regions $A_{i}$ are joined in this way, all reduced circular diagrams simplify to just two equivalent joint circular diagrams, as shown in Fig.\ref{fig-Q52}(i).

This motivates us to prove the joint form of a given HEIs, rather than their original form, provided that their circular diagrams are gapless.

There are two important features of the joint form:

\begin{enumerate}
\item Complementarity: if $\bar{K}$ is a complement of region $K$, i.e. $\bar{K}=\left\{
1,...,n\right\}  / K$, then the corresponding entropies are equal,
\begin{equation}
S_{\bar{K}}=S_{K}.
\end{equation}
This identity allows one to simplify a HEI by eliminating one region and replacing its complementary counterpart.

\item Dual representation: any RT surface can be expressed equivalently in two complementary simplex notation, e.g.
\begin{equation}
S_{\left[  ij\right]  }=S_{\left[  j+1,i-1\right]  }.
\end{equation}
This relation becomes particularly useful when gaps between subsystems are closed.
\end{enumerate}

Consider the 5-partite inequality $Q_{2}^{(5)}$ in eq.(\ref{Q52}). Because its circular diagram is gapless for all CCC configurations, we can close the gap corresponding to region 5 and remove $A_5$ using the complementarity relations: 
\begin{align}
&S_{25}=S_{134},S_{45}=S_{123},S_{125}=S_{34},S_{145}=S_{23},\\
&S_{235}=S_{14},S_{245}=S_{13},S_{1235}=S_{4},S_{1245}=S_{3}.
\end{align}
Substituting these identities into eq.(\ref{Q52}), we obtain the 4-partite joint form: 
\begin{equation}
S_{123}+S_{124}+S_{34}\geq S_{12}+S_{1234}+S_{3}+S_{4},\label{Q52 adj}%
\end{equation}
which involves only the four regions $A_1$, $A_2$, $A_3$, $A_4$.

For the two CCC configurations of the 4-partite system (with either $S_{13}^{d}$ or  $S_{24}^{d}$ disconnected), eq.(\ref{Q52 adj}) reduces in the simplex basis to
\begin{equation}
S_{\left[  13\right]  }+S_{\left[  34\right]  }\geq S_{\left[  12\right]
}+S_{\left[  44\right]  },
\end{equation}
which corresponds to the right-hand diagram in  in Fig.\ref{fig-Q52}(i). The inequality follows directly from the cross inequality. Hence, the joint form eq.(\ref{Q52 adj}) is valid, and consequently the original 5-partite inequality $Q_{2}^{(5)}\ge 0$ is proven.

Now we have an improved strategy to prove a HEI in a CCC configuration:

\begin{enumerate}
\item Reduce the HEI to its joint form.

\item Express the the joint form in the simplex basis.

\item Construct the joint circular diagrams and prove the inequality by using cross inequalities.
\end{enumerate}

Before applying this approach to other HEIs, it is useful to recall a general reduction identity in the I-basis,
\begin{equation}
I_{Ki}=I_{K}+\sum_{Q\subseteq\left(  \bar{K}/i\right)  }\left(  -1\right)
^{\left\vert KQ\right\vert +1}I_{KQ},\label{IKi}%
\end{equation}
which systematically shows how to remove a selected region $i$ for an I-basis.

Applying the improved strategy, one can quickly verify the remaining  three HEIs
$Q_{3}^{\left(  5\right)  }$, $Q_{4}^{\left(  5\right)  }$ and $Q_{5}^{\left(5\right)  }$ as follows:
\begin{equation}
Q_{3}^{\left(  5\right)  }=-I_{125}-I_{135}-I_{145}-I_{234}+I_{1235}%
+I_{1245}+I_{1345}\geq0
\label{Q53}
\end{equation}
reduces via eq.(\ref{IKi}) to
\begin{equation}
Q_{3}^{\left(  5\right)  }=-I_{234}\geq0,
\end{equation}
which is simply a MMI for the regions $A_2$, $A_3$, $A_4$.
\begin{equation}
Q_{4}^{\left(  5\right)  }=-I_{123}-I_{145}-I_{234}-I_{235}+I_{1234}%
+I_{1235}\geq0
\end{equation}
reduces to
\begin{equation}
Q_{4}^{\left(  5\right)  }=-I_{124}-I_{134}-2I_{234}+2I_{1234}\geq
0.\label{Q53}%
\end{equation}
In the two CCC configurations with $S_{13}^{d}$ and $S_{24}^{d}$ disconnected, the inequality becomes
\begin{align}
S_{13}^{d} &  :2S_{\left[  13\right]  }+S_{\left[  24\right]  }+S_{\left[
34\right]  }\geq S_{\left[  11\right]  }+S_{\left[  12\right]  }+2S_{\left[
44\right]  },\\
S_{24}^{d} &  :S_{\left[  13\right]  }+S_{\left[  34\right]  }\geq S_{\left[
12\right]  }+S_{\left[  44\right]  },
\end{align}
both of which follow immediately from cross inequalities.
\begin{equation}
Q_{5}^{\left(  5\right)  }=-I_{123}-2I_{125}-2I_{134}-I_{145}-I_{234}%
-I_{235}+2I_{1234}+2I_{1235}+I_{1245}+I_{1345}\geq0,
\end{equation}
reduces to
\begin{equation}
Q_{3}^{\left(  5\right)  }=-I_{124}-I_{134}-2I_{234}+2I_{1234}\geq0,
\end{equation}
which is the same as the joint form of the HEI $Q_{4}^{\left(  5\right)  }$.

In conclusion, proving an HEI through its joint form is considerably simpler than verifying all original CCC configurations individually.
The equivalence between the joint and original forms holds whenever the circular diagrams are gapless. This observation leads naturally to the  following Gapless Theorem.

\noindent\textbf{Gapless Theorem}: a superbalanced HEI in a CCC configuration
is always gapless.

\noindent\textbf{Proof}: For a $n$-partite information $I_{k_{1}\cdots k_{n}}%
$\ with $n\geq3$, every connected RT surface
\begin{equation}
S_{{k_{1}}...k_{i}k_{i+1}\cdots{k_{m}}}^{c}=\sum_{\sigma=1}^{m}S_{\left[
{k_{\sigma+1}k_{\sigma}}\right]  },
\end{equation}
contributes a term $S_{\left[  k_{i+1}k_{i}\right]  }$. So, the number of terms $S_{\left[
k_{i+1}k_{i}\right]  }$ in a completely connected configuration is
\begin{equation}
\sum_{m=0}^{n-1}\left(  -1\right)  ^{m+1}C_{m}^{n-2}=0.
\end{equation}
On the other hand, to avoid incompatible pairs, we should apply for certain cuts to obtain CCC configurations. However, these cuts do not include ones in the form $C_{j+1,j}^{j,j+1}$ so that the number of terms $S_{\left[  j+1,j\right]  }$ remains zero in the CCC configurations, i.e. a superbalanced HEI in a CCC configuration is always gapless. $\Box$

\subsection{HEIs in Non-CCC Configurations}

We have constructed a systematic method to prove HEIs in the CCC
configurations. In this section, we will show that a HEI holding in the CCC
configurations ensures that it holds in all other configurations.

Starting from a CCC configuration, we can obtain another configuration by cutting
a connected term. Through repeated cuts, all configurations can be obtained. We
are going to show that applying a cut will not affect the validity of HEIs.

First, let us formally introduce a concept called a cut $C_{kl}^{ij}$, which is a
segment that intersects with the two RT surfaces $S_{\left[  ij\right]  }$ and
$S_{\left[  kl\right]  }$ in a connected HEE. The result of applying a cut
$C_{kl}^{ij}$ in a connected HEE is that the pair of lines $S_{\left[
ij\right]  }$ and $S_{\left[  kl\right]  }$ break and rejoin to a new pair of
lines, so that the original connected HEE will divide into two connected HEEs
by a cut.

\begin{figure}[ptb]
\centering{ \subfloat[$S_{1234}^{c}(C_{32}^{14})=
S_{12}^{c}+S_{34}^{c}$]{\includegraphics[width=.23\linewidth]{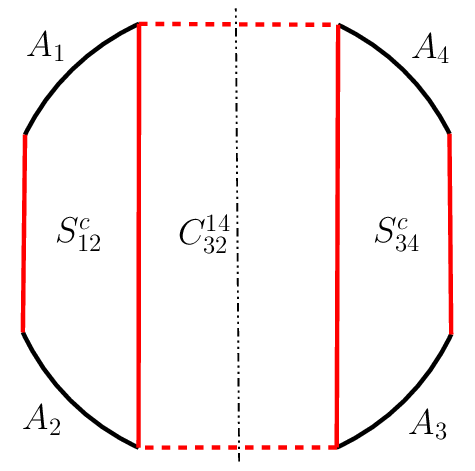} }\hspace
{.12cm} \subfloat[$S_{123}^{c}(C_{32}^{13})=
S_{12}^{c}+S_{3}$]{\includegraphics[width=.23\linewidth]{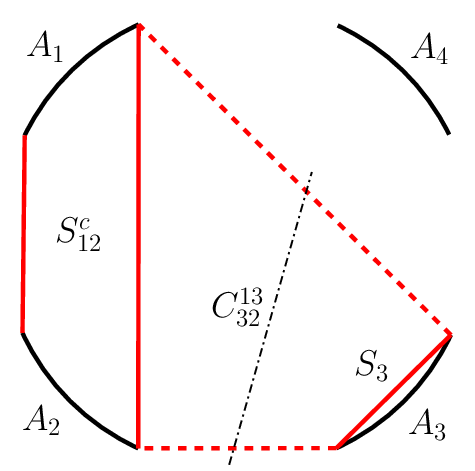} }\hspace
{.12cm}
\subfloat[$C_{32}^{14}$]{\includegraphics[width=.23\linewidth]{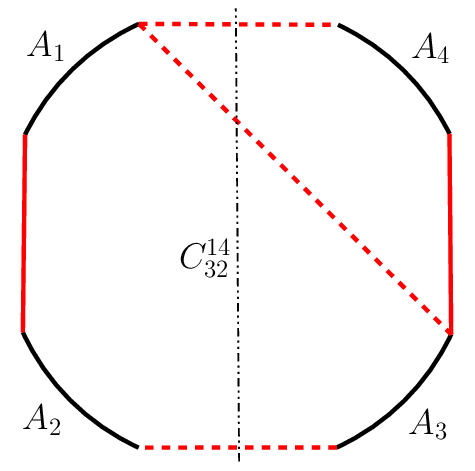} }\hspace
{.12cm}
\subfloat[$C_{32}^{21}$]{\includegraphics[width=.23\linewidth]{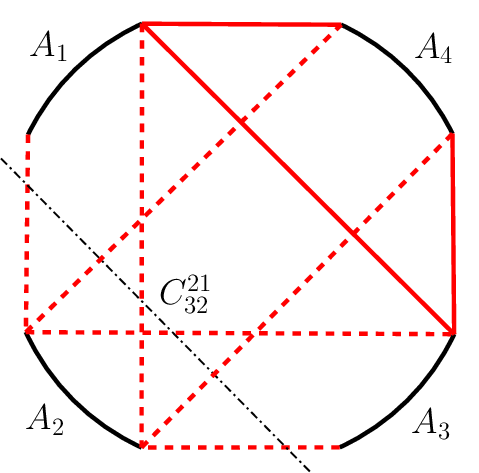} }\hspace
{.12cm}}\caption{The splits of $S_{1234}^{c}$ and $S_{123}^{c}$}%
\label{fig-1234-split}%
\end{figure}

We will use the following examples in a 4-partite system to illustrate the
effect of a cut as shown in Fig.\ref{fig-1234-split}. We first consider
$S_{1234}^{c}=S_{\left[  14\right]  }+S_{\left[  43\right]  }+S_{\left[
32\right]  }+S_{\left[  21\right]  }$. In Fig.\ref{fig-1234-split}(a), we
apply a cut $C_{32}^{14}$ that breaks the two dashed lines $S_{\left[
32\right]  }$ and $S_{\left[  14\right]  }$, and rejoins them to the two solid
lines $S_{\left[  12\right]  }$ and $S_{\left[  34\right]  }$. The result is
that $S_{1234}^{c}$ splits into two lines $S_{12}^{c}=S_{\left[  12\right]
}+S_{\left[  21\right]  }$ and $S_{34}^{c}=S_{\left[  34\right]  }+S_{\left[
43\right]  }$, i.e. $S_{1234}^{c}\left(  C_{32}^{14}\right)  =S_{12}%
^{c}+S_{34}^{c}$.

Fig.\ref{fig-1234-split}(b) shows another example of splitting $S_{123}%
^{c}=S_{\left[  13\right]  }+S_{\left[  32\right]  }+S_{\left[  21\right]  }$
by applying a cut $C_{32}^{13}$, which breaks the two dashed lines $S_{\left[
13\right]  }$ and $S_{\left[  32\right]  }$, and rejoins them to the two solid
lines $S_{\left[  12\right]  }$ and $S_{\left[  33\right]  }$. Thus
$S_{123}^{c}$ splits into $S_{12}^{c}=S_{\left[  12\right]  }+S_{\left[
21\right]  }$ and $S_{3}=S_{\left[  33\right]  }$, i.e. $S_{123}^{c}\left(
C_{32}^{13}\right)  =S_{12}^{c}+S_{3}$.

In Fig.\ref{fig-1234-split}(c), we mix the above two examples of $S_{1234}%
^{c}$\ and $S_{123}^{c}$ together and apply the cut $C_{32}^{14}$. We observe
that the cut $C_{32}^{14}$ splits not only $S_{1234}^{c}$ but also
$S_{123}^{c}$ simultaneously, i.e. $S_{1234}^{c}\left(  C_{32}^{14}\right)
=S_{12}^{c}+S_{34}^{c}$ and $S_{123}^{c}\left(  C_{32}^{14}\right)
=S_{12}^{c}+S_{3}$. In addition, we notice that both cuts $C_{32}^{13}$ and
$C_{32}^{14}$ split $S_{123}^{c}$ into $S_{12}^{c}+S_{3}$.

Our last example in Fig.\ref{fig-1234-split}(d) shows that the cut
$C_{32}^{21}$ splits $S_{12}^{c}\left(  C_{32}^{21}\right)  =S_{1}+S_{2}$,
$S_{23}^{c}\left(  C_{32}^{21}\right)  =S_{2}+S_{3}$, $S_{24}^{c}\left(
C_{32}^{21}\right)  =S_{2}+S_{4}$, $S_{123}^{c}\left(  C_{32}^{21}\right)
=S_{2}+S_{13}^{c}$, $S_{124}^{c}\left(  C_{32}^{21}\right)  =S_{2}+S_{14}^{c}%
$, $S_{234}^{c}\left(  C_{32}^{21}\right)  =S_{2}+S_{34}^{c}$ and
$S_{1234}^{c}\left(  C_{32}^{21}\right)  =S_{2}+S_{134}^{c}$ simultaneously.
In other words, the cut $C_{32}^{21}$ decouples $A_{2}$ from the other
entanglement regions.

These observations can be summarized into the following Cut Theorem.

\begin{figure}[ptb]
\centering{ \includegraphics[width=.3\linewidth]{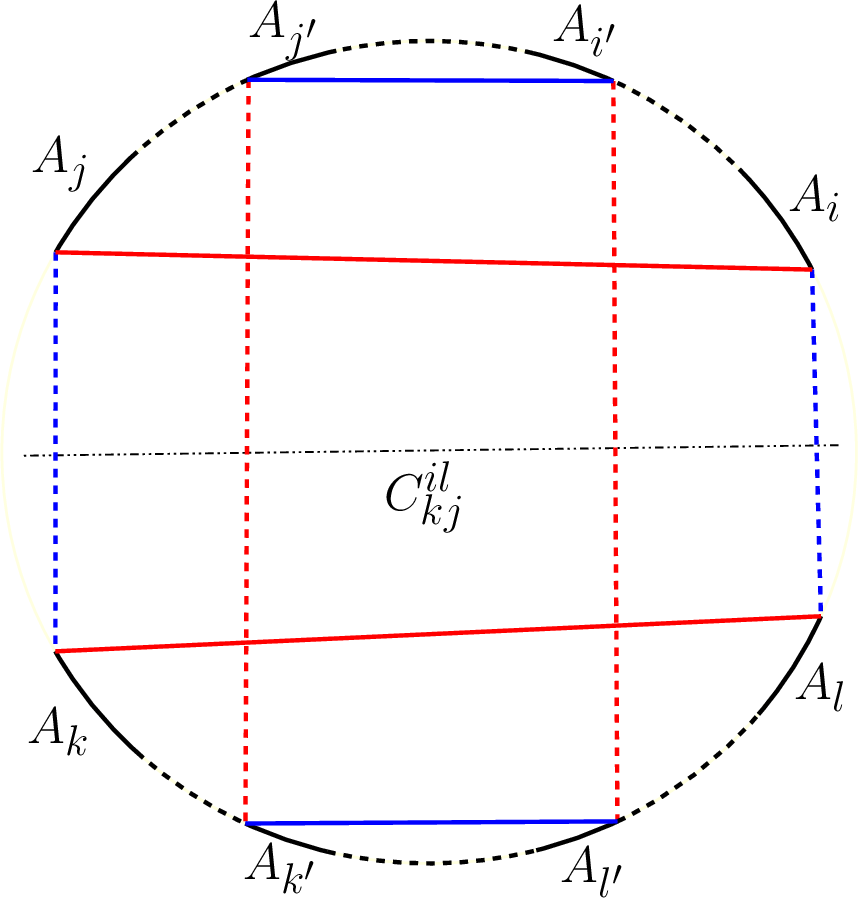} } \caption{The circular diagram of eq.(\ref{split}).}%
\label{fig-cut}%
\end{figure}

\noindent\textbf{Cut Theorem}: A cut $C_{kj}^{il}$\ splits $S_{{i}^{\prime
}{\cdots j}^{\prime}{k}^{\prime}{\cdots l}^{\prime}}^{c}$ into $S_{{i}%
^{\prime}{\cdots j}^{\prime}}^{c}$ and $S_{{k}^{\prime}{\cdots l}^{\prime}%
}^{c}$, i.e. $S_{{i}^{\prime}{\cdots j}^{\prime}{k}^{\prime}{\cdots l}%
^{\prime}}^{c}\left(  C_{kj}^{il}\right)  =S_{{i}^{\prime}{\cdots j}^{\prime}%
}^{c}+S_{{k}^{\prime}{\cdots l}^{\prime}}^{c}$, if $\left[  {i}^{\prime}%
,{j}^{\prime}\right]  \subseteq\left[  {i,j}\right]  $ and $\left[
{k}^{\prime}{,l}^{\prime}\right]  \subseteq\left[  {k,l}\right]  $, i.e.
$C_{kj}^{il}$ induces $C_{k^{\prime}j^{\prime}}^{i^{\prime}l^{\prime}}$.

\noindent\textbf{Proof}: By the definition of a cut,%
\begin{equation}
S_{{i\cdots jk\cdots l}}^{c}\left(  C_{kj}^{il}\right)  \rightarrow
S_{{i\cdots j}}^{c}+S_{{k\cdots l}}^{c},
\end{equation}
which implies%
\begin{equation}
S_{{i\cdots jk\cdots l}}^{c}\geq S_{{i\cdots j}}^{c}+S_{{k\cdots l}}^{c}.
\end{equation}
Thus
\begin{equation}
S_{[il]}+S_{[kj]}+\cdots\geq S_{[{ij}]}+S_{[{kl}]}+\cdots,\nonumber
\end{equation}
where $\ldots$ on both sides are the same. Then, we have%
\begin{equation}
S_{[il]}+S_{[kj]}-S_{[{ij}]}-S_{[{kl}]}\geq0. \label{Sk}%
\end{equation}
To show%
\begin{equation}
S_{{i}^{\prime}{\cdots j}^{\prime}{k}^{\prime}{\cdots l}^{\prime}}^{c}\left(
C_{kj}^{il}\right)  \rightarrow S_{{i}^{\prime}{\cdots j}^{\prime}}^{c}%
+S_{{k}^{\prime}{\cdots l}^{\prime}}^{c},
\end{equation}
we need to prove%
\begin{equation}
S_{[i^{\prime}l^{\prime}]}+S_{[k^{\prime}j^{\prime}]}-S_{[{i}^{\prime}%
{j}^{\prime}]}-S_{[{k}^{\prime}{l}^{\prime}]}\geq0.
\end{equation}
It is sufficient to show that%
\begin{equation}
S_{[i^{\prime}l^{\prime}]}+S_{[k^{\prime}j^{\prime}]}-S_{[{i}^{\prime}%
{j}^{\prime}]}-S_{[{k}^{\prime}{l}^{\prime}]}\geq S_{[il]}+S_{[kj]}-S_{[{ij}%
]}-S_{[{kl}]}\nonumber
\end{equation}
or
\begin{equation}
S_{[i^{\prime}l^{\prime}]}+S_{[k^{\prime}j^{\prime}]}+S_{[{ij}]}+S_{[{kl}%
]}\geq S_{[il]}+S_{[kj]}+S_{[{i}^{\prime}{j}^{\prime}]}+S_{[{k}^{\prime}%
{l}^{\prime}]}. \label{split}%
\end{equation}
Therefore, we need to show that the sum of the red lines is greater than the
sum of the blue lines illustrated in Fig.\ref{fig-cut}, which can be verified by using the
clean-gap procedure. This completes the proof of the Cut Theorem. $\Box$

We have shown that a connected RT surface can be split into two connected
ones by applying a cut. This step can be performed repeatedly from a HEI in a CCC
configuration. All possible configurations can be obtained by applying a finite
number of cuts.

For a $n$-partite system, there are at most $\frac{n\left(  n+1\right)  }{2}$
lines in a CCC configuration. For a line $\left[  ij\right]  $, there are
$\sum_{k=1}^{j-i}k$ dual segments to form non-intersectred pairs. Thus, there
are in total%
\begin{equation}
N_{n}=\frac{n}{2}\sum_{m=1}^{n-1}\sum_{k=1}^{m}k=\frac{n^{2}\left(
n^{2}-1\right)  }{12}
\end{equation}
non-intersected pairs, and each pair is associated to a cut.

Among all the cuts, some cuts will induce other ones. We therefore classify the
cuts into $\left(  n-1\right)  $ levels for a $n$-partite system. There are
$\frac{nk\left(  n-k\right)  }{2}$ cuts in the $k$-level and each of them
induces $\left(  k-1\right)  $-level cuts and therefore the lower level cuts.

\begin{itemize}
\item at level $k=1$, there are $\frac{n\left(  n-1\right)  }{2}$ lowest cuts
$C_{ji}^{ij}$: a cut that does not induce any other cut.

\item at level $k=n-1$ , there are $\frac{n\left(  n-1\right)  }{2}$ highest
cuts: a cut that completely divides the circular diagram into two independent parts.
\end{itemize}

Let's list the cuts for some lower $n$-partite systems.

\begin{figure}[ptb]
\centering{
\subfloat[3-partite cuts DAG]{\includegraphics[width=.2\linewidth]{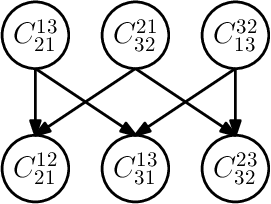} }\hspace
{2cm}
\subfloat[4-partite cuts DAG]{\includegraphics[width=.5\linewidth]{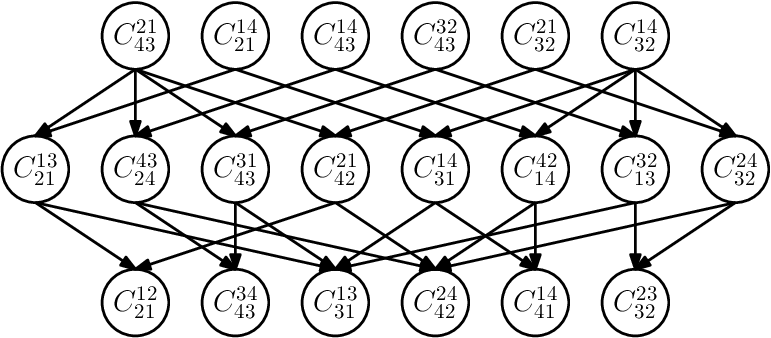} }\hspace
{.12cm}
\subfloat[5-partite cuts DAG]{\includegraphics[width=.9\linewidth]{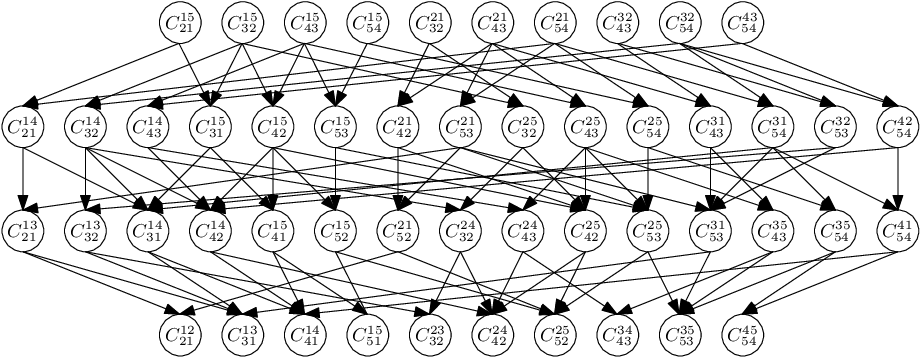} }\hspace
{.12cm}} \caption{The Dependency Acyclic Graph (DAG) for cuts.}%
\label{cut flow}%
\end{figure}

\begin{itemize}
\item For a tripartite system, there are in total $N_{3}=\frac{3^{2}\left(
3^{2}-1\right)  }{12}=6$ cuts. They are $3$ level 1 (lowest) cuts and $3$
level 2 (highest) cuts which induce the level 1 cuts as shown in Fig.\ref{cut flow}(a).

\item For a 4-partite system, there are in total $N_{4}=\frac{4^{2}\left(
4^{2}-1\right)  }{12}=20$ cuts. They are $6$ level 1 (lowest) cuts, $8$ level
2 cuts and $6$ level 3 (highest) cuts as shown in Fig.\ref{cut flow}(b).

\item For a 5-partite system, there are in total $N_{5}=\frac{5^{2}\left(
5^{2}-1\right)  }{12}=50$ cuts. They are $10$ level 1 (lowest) cuts, $15$
level 2 cuts, $15$ level 3 cuts and $10$ level 6 (highest) cuts as shown in
Fig.\ref{cut flow}(c).
\end{itemize}

Since applying any combination of cuts on a CCC configuration leads to a new
configuration, naively, we need to consider in total $2^{N_{n}}$ configurations
for a $n$-partite system. This number grows rapidly with the number of
entanglement regions $n$ and it is usually impossible to check them one by
one. However, since higher level cuts induce lower level cuts, many of the
naive configurations are not allowed.

To count the number of allowed configurations, we realize that the cuts
form a Dependency Acyclic Graph (DAG) for any $n$-partite system. We can use
dynamical programming methods to count the independent paths of a DAG, and
each independent path is associated with an allowed configuration.

For a tripartite system, there are naively $2^{6}=64$ configurations. Among
them, $17$ configurations are allowed. For a $4$-partite system, there are
naively $2^{20}=1048576$ configurations with $1570$ of them being allowed
(after considering compatibility). For a $5$-partite system, there are naively
$2^{50}=1125899906842624$ configurations with $2864048$ of them being allowed
(after considering compatibility).

We see that the number of allowed configurations is much less than the number
of naive configurations. Nevertheless, checking all allowed configurations
one by one is still a tedious job. Remarkably, we can prove that if a HEI
holds in CCC configurations, then it holds in any other configurations.

\noindent\textbf{Configuration Theorem}. If a balanced HEI is valid in CCC
configurations, then it is valid in all configurations.

\noindent\textbf{Proof}. Any configuration can be obtained from a CCC
configuration by a sequence of cuts. In addition, a superbalanced HEI can be
reduced to a balanced HEI by applying a cut. Therefore, by mathematical
induction, it suffices to prove the following statement:

If a balanced HEI is valid in a configuration $C$, which is not necessarily
a CCC configuration, then the reduced HEI obtained by applying a cut
is also valid in the same configuration $C$ with that cut.

However, it is straightforward to show that the reduced HEI in simplex
basis has the same form for the original configuration $C$ with or without the cut. Consequently, we are free to prove the validity of the reduced
HEI either in the configuration $C$ or in the configuration $C$ with the cut.
This is the key observation used to establish the Configuration Theorem.

\begin{enumerate}
\item Assume that a balanced HEI
\begin{equation}
Q_{n}^{b}=\sum_{\left\vert K\right\vert \geq 2} q_{K} I_{K} \geq 0,
\end{equation}
is valid in a configuration $C$.

\item Apply a cut $C_{kj}^{il}$ to this HEI. The reduced HEI is
\begin{equation}
Q_{n}^{b}\!\left(C_{kj}^{il}\right)
= \sum_{\left\vert K\right\vert \geq 2} q_{K} I_{K}\!\left(C_{kj}^{il}\right),
\end{equation}
where the action of the cut $C_{kj}^{il}$ on the $I$-basis can be calculated as
\begin{equation}
I_{K}\!\left(C_{kj}^{il}\right)
= I_{K}
+ \sum_{\substack{I\subseteq K\cap\left[{i,j}\right]\\[2pt]
J\subseteq K\cap\left[{k,l}\right]}}
(-1)^{\left\vert IJ\right\vert+1} I_{(I)(J)}.
\end{equation}
Therefore,
\begin{align}
Q_{n}^{b}\!\left(C_{kj}^{il}\right)
&= \sum_{\left\vert K\right\vert \geq 2} q_{K} I_{K}
+ \sum_{\left\vert K\right\vert \geq 2} q_{K}
\sum_{\substack{I\subseteq K\cap\left[{i,j}\right]\\
J\subseteq K\cap\left[{k,l}\right]}}
(-1)^{\left\vert IJ\right\vert+1} I_{(I)(J)} \nonumber\\
&= Q_{n}^{b}
+ \sum_{\substack{I\subseteq\left[{i,j}\right]\\
J\subseteq\left[{k,l}\right]}}
\biggl[\,(-1)^{\left\vert IJ\right\vert+1}
\sum_{K^{\prime}\subseteq\overline{\left[{i,j}\right]\cup\left[{k,l}\right]}}
q_{IJK^{\prime}} \biggr] I_{(I)(J)} \nonumber\\
&= Q_{n}^{b}
+ \sum_{\substack{I\subseteq\left[{i,j}\right]\\
J\subseteq\left[{k,l}\right]}}
q_{IJ}^{\prime}\, I_{(I)(J)},
\label{Q-cut}
\end{align}
where
\begin{equation}
q_{IJ}^{\prime}
= (-1)^{\left\vert IJ\right\vert+1}
\sum_{K^{\prime}\subseteq\overline{\left[{i,j}\right]\cup\left[{k,l}\right]}}
q_{IJK^{\prime}}.
\end{equation}

Now consider evaluating the reduced HEI $Q_{n}^{b}\!\left(C_{kj}^{il}\right)$
in the configuration $C$ with or without the cut
$C_{kj}^{il}$. By assumption, the first term in eq.\eqref{Q-cut},
$Q_{n}^{b}$, is strictly greater than $0$ in the configuration $C$.
For the remaining terms, we distinguish two cases:

\begin{enumerate}
\item For those terms whose coefficients satisfy $q_{IJ}^{\prime} \geq 0$, we
choose to evaluate them in the configuration $C$ without the cut $C_{kj}^{il}$.
In this case, we have $S_{I} + S_{J} \geq S_{IJ}^{c}$, which implies
\begin{equation}
I_{(I)(J)} = S_{I} + S_{J} - S_{IJ}^{c} \geq 0,
\end{equation}
and hence
$q_{IJ}^{\prime} I_{(I)(J)} \geq 0$. Therefore, the sum of the two
nonnegative quantities is nonnegative:
\begin{equation}
Q_{n}^{b} + q_{IJ}^{\prime} I_{(I)(J)} \geq 0.
\end{equation}

\item For those terms whose coefficients satisfy $q_{IJ}^{\prime} \leq 0$, we
choose to evaluate them in the configuration $C$ with the cut $C_{kj}^{il}$.
In that configuration, we have $S_{I} + S_{J} \leq S_{IJ}^{c}$, which implies
\begin{equation}
I_{(I)(J)} = S_{I} + S_{J} - S_{IJ}^{c} \leq 0,
\end{equation}
and thus again
$q_{IJ}^{\prime} I_{(I)(J)} \geq 0$. Consequently, the sum of the two
nonnegative quantities is nonnegative:
\begin{equation}
Q_{n}^{b} + q_{IJ}^{\prime} I_{(I)(J)}\!\left(C_{kj}^{il}\right) \geq 0.
\end{equation}

\end{enumerate}
\end{enumerate}

This completes the proof of the Configuration Theorem. $\Box$

\begin{figure}[ptb]
\centering{
\subfloat[]{\includegraphics[width=.25\linewidth]{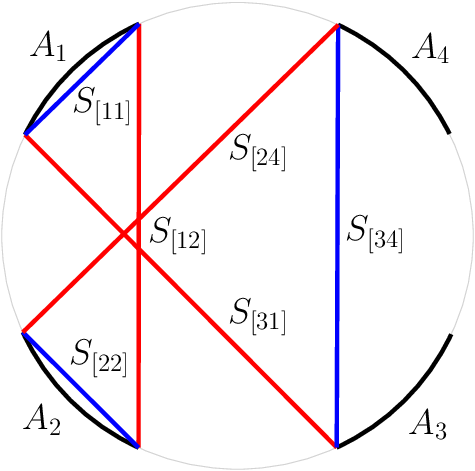}}\hspace{0.5cm}
\subfloat[]{\includegraphics[width=.25\linewidth]{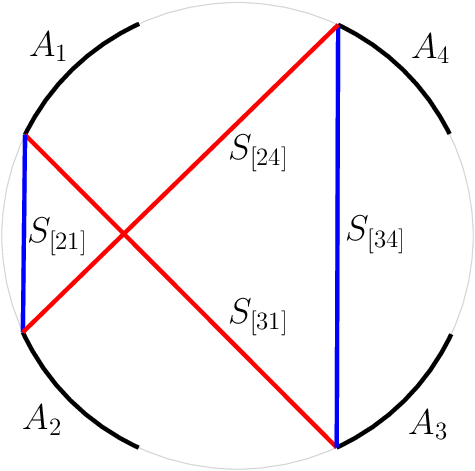}}\hspace{0.5cm}
\subfloat[]{\includegraphics[width=.25\linewidth]{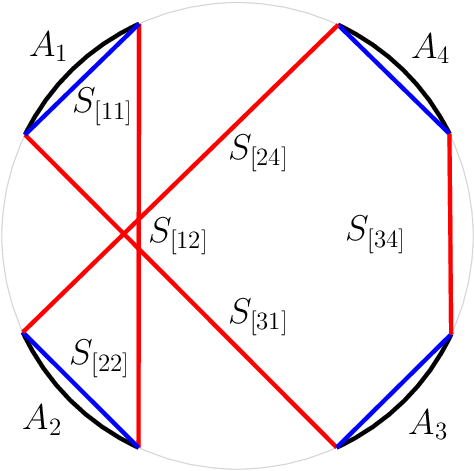}}}
\caption{(a) Circular diagram corresponding to eq.\eqref{exICCC}.
(b) Circular diagram corresponding to eq.\eqref{exI122}.
(c) Circular diagram corresponding to eq.\eqref{exI342}.}%
\label{exI}%
\end{figure}

The arguments in the above proof are quite formal. In this subsection, we
illustrate the Configuration Theorem with a concrete example. Consider the
superbalanced HEI in a 4-partite system
\begin{equation}
Q = -I_{123} - I_{124} + I_{1234} \geq 0. \label{ex}%
\end{equation}
In a 4-partite system, there are two CCC configurations,
$\{ S_{1,3}^{d} \}$ and $\{ S_{2,4}^{d} \}$. Since these two CCC
configurations are related by a symmetry, it suffices to consider only one of
them, say $\{ S_{1,3}^{d} \}$. Expressing the HEI in eq.\eqref{ex} in
simplex basis of this CCC configuration, we obtain
\begin{equation}
Q = -S_{[1,1]} + S_{[1,2]} - S_{[2,2]} + S_{[2,4]} + S_{[3,1]} - S_{[3,4]},
\label{exICCC}%
\end{equation}
whose circular diagram is shown in Fig.\ref{exI}(a). It is straightforward to
verify that $Q > 0$ by using cross inequalities.

Next, we apply a cut $C_{21}^{12}$ to $Q$:
\begin{equation}
Q\!\left(C_{21}^{12}\right)
= -I_{123} - I_{124} + I_{1234} + I_{12}
= Q + I_{12}, \label{exI12}%
\end{equation}
which, in simplex basis becomes
\begin{equation}
Q\!\left(C_{21}^{12}\right)
= -S_{[2,1]} + S_{[2,4]} + S_{[3,1]} - S_{[3,4]},
\label{exI122}
\end{equation}
and its circular diagram is plotted in Fig.\ref{exI}(b).
As in the case of $Q$, it is straightforward to show that
$Q\!\left(C_{21}^{12}\right) > 0$ by a cross inequality.

Now consider applying a different cut, $C_{43}^{34}$, to $Q$:
\begin{equation}
Q\!\left(C_{43}^{34}\right)
= -I_{123} - I_{124} + I_{1234} - I_{34}
= Q - I_{34}. \label{exI34}%
\end{equation}
In this case, the coefficient of $I_{34}$ is negative.

In the simplex basis, eq.\eqref{exI34} becomes
\begin{equation}
Q\!\left(C_{43}^{34}\right)
= -S_{[1,1]} + S_{[1,2]} - S_{[2,2]} + S_{[2,4]}
+ S_{[3,1]} - S_{[3,3]} + S_{[4,3]} - S_{[4,4]},
\label{exI342}
\end{equation}
whose circular diagram is plotted in Fig.\ref{exI}(c). As expected, it is
not possible to show that $Q\!\left(C_{43}^{34}\right) > 0$ using only cross
inequalities. To establish its positivity, we must use the additional
information that $C_{43}^{34}$ cuts $S_{34} = S_{3} + S_{4}$, i.e.
\begin{equation}
S_{34}^{c} \geq S_{3} + S_{4},
\end{equation}
which implies
\begin{equation}
S_{[3,4]} + S_{[4,3]} \geq S_{[3,3]} + S_{[4,4]}.
\end{equation}
Therefore,
\begin{equation}
Q\!\left(C_{43}^{34}\right)
\geq -S_{[1,1]} + S_{[1,2]} - S_{[2,2]} + S_{[2,4]}
+ S_{[3,1]} - S_{[3,4]},
\end{equation}
which is exactly the same expression as in eq.\eqref{exICCC}. Since
eq.\eqref{exICCC} has already been shown to be strictly positive, we conclude
that
\begin{equation}
Q\!\left(C_{43}^{34}\right) \geq 0
\end{equation}
holds conditionally, due to the constraint imposed by the cut $C_{43}^{34}$.

We have thus shown that a balanced HEI which is valid in certain configurations
remains (conditionally) valid after applying a cut. We are now
ready to present a complete strategy for proving a superbalanced HEI:

\begin{enumerate}
\item Obtain the joint form of the superbalanced HEI.

\item Express this joint form in CCC configurations using the simplex basis.

\item Plot the corresponding joint circular diagram and prove its validity by
      cross inequalities.

\item Conclude that the HEI is valid in all configurations, by the
      Configuration Theorem.
\end{enumerate}

\section{Other Examples}

In this section, we present further examples for $n=6$ and $7$.

In \cite{2309.06296}, a family of HEIs in the 6-partite system was discovered. We will prove two of them using our strategy as illustrative examples. Using
the notation of \cite{2309.06296}, the two HEEs we consider are
\begin{align}
Q_{\left\{  4,3\right\}  }^{\left[  24\right]  } &  =-I_{123}-I_{126}
-I_{156}-I_{246}+I_{1236}+I_{1246}+I_{1256},\\
Q_{\left\{  5,3\right\}  }^{\left[  13\right]  } &  =-I_{123}-I_{125}
-I_{126}-I_{356}-I_{456}+I_{1236}+I_{1256}+I_{3456}.
\end{align}

By eliminating the region $A_{6}$, the above two HEIs reduce to their joint
forms
\begin{align}
Q_{\left\{  4,3\right\}  }^{\left[  24\right]  } &  =-I_{135}-I_{145}%
-I_{234}-I_{245}+I_{1245}+I_{1345}+I_{2345},\\
Q_{\left\{  5,3\right\}  }^{\left[  13\right]  } &  =-I_{124}-I_{135}%
-I_{145}-I_{235}-I_{245}+I_{1245}+I_{1345}+I_{2345},
\end{align}
which is effectively a HEI in a 5-partite system. We now study these joint forms in the $11$ CCC configurations for $n=5$.

For $Q_{\left\{  4,3\right\}  }^{\left[  24\right]  }$, in $11$ CCC configurations
we obtain
\begin{align}
\text{CCC}1,4 &  :2S_{\left[  14\right]  }+S_{\left[  23\right]  }+S_{\left[
25\right]  }+2S_{\left[  35\right]  }\geq S_{\left[  11\right]  }+S_{\left[
12\right]  }+S_{\left[  22\right]  }+S_{\left[  45\right]  }+2S_{\left[
55\right]  }\\
\text{CCC}2 &  :S_{\left[  14\right]  }+S_{\left[  23\right]  }+S_{\left[
25\right]  }+S_{\left[  35\right]  }\geq S_{\left[  11\right]  }+S_{\left[
22\right]  }+S_{\left[  45\right]  }+S_{\left[  55\right]  }\\
\text{CCC}3,5,8,9 &  :S_{\left[  14\right]  }+S_{\left[  23\right]
}+2S_{\left[  35\right]  }\geq S_{\left[  12\right]  }+S_{\left[  22\right]
}+S_{\left[  45\right]  }+S_{\left[  55\right]  }\\
\text{CCC}6 &  :2S_{\left[  14\right]  }+S_{\left[  23\right]  }+2S_{\left[
35\right]  }\geq S_{\left[  11\right]  }+S_{\left[  12\right]  }+S_{\left[
33\right]  }+2S_{\left[  55\right]  }\\
\text{CCC}7 &  :S_{\left[  23\right]  }+S_{\left[  35\right]  }\geq S_{\left[
22\right]  }+S_{\left[  45\right]  }\\
\text{CCC}10 &  :S_{\left[  14\right]  }+S_{\left[  23\right]  }+S_{\left[
35\right]  }\geq S_{\left[  11\right]  }+S_{\left[  33\right]  }+S_{\left[
55\right]  }\\
\text{CCC}11 &  :S_{\left[  14\right]  }+S_{\left[  23\right]  }+2S_{\left[
35\right]  }\geq S_{\left[  12\right]  }+S_{\left[  25\right]  }+S_{\left[
33\right]  }+S_{\left[  55\right]  }%
\end{align}
each of which can be verified straightforwardly by cross inequalities.

For $Q_{\left\{  5,3\right\}  }^{\left[  13\right]  }$, in $11$ CCC configurations
we obtain,
\begin{align}
\text{CCC}1,2,4,5,9 &  :S_{\left[  14\right]  }+S_{\left[  24\right]
}+2S_{\left[  35\right]  }+S_{\left[  41\right]  }+S_{\left[  52\right]  }\geq
S_{\left[  11\right]  }+S_{\left[  22\right]  }+2S_{\left[  34\right]
}+S_{\left[  45\right]  }+S_{\left[  55\right]  }\\
\text{CCC}3,7,8 &  :S_{\left[  24\right]  }+2S_{\left[  35\right]
}+S_{\left[  41\right]  }+S_{\left[  52\right]  }\geq S_{\left[  22\right]
}+S_{\left[  31\right]  }+S_{\left[  34\right]  }+S_{\left[  45\right]
}+S_{\left[  55\right]  }\\
\text{CCC}6,10,11 &  :S_{\left[  14\right]  }+S_{\left[  24\right]
}+2S_{\left[  35\right]  }+S_{\left[  41\right]  }+S_{\left[  52\right]  }\geq
S_{\left[  11\right]  }+S_{\left[  25\right]  }+2S_{\left[  34\right]
}+S_{\left[  42\right]  }+S_{\left[  55\right]  }%
\end{align}
which again can be proved directly by cross inequalities.

Finally, we turn to the $7$-partite system. In \cite{2309.06296}, the following HEI in $7$-partite system was discovered,
\begin{align}
&  S_{1245}+S_{1246}+S_{1257}+S_{1456}+S_{1457}+S_{1345}+S_{1346}\nonumber\\
+&S_{1357}+S_{2456}+S_{2457}+S_{2345}+S_{2346}+S_{2357}+S_{3456}+S_{3457}%
\nonumber\\
 \geq & S_{123}+S_{145}+S_{146}+S_{157}+S_{245}+S_{246}+S_{257}+S_{345}\nonumber\\
+&S_{346}+S_{357}+S_{12456}+S_{12457}+S_{13456}+S_{13457}+S_{23456}+S_{23457},
\end{align}
which can be written in the superbalanced form using I-basis as
\begin{align}
Q_{7} &  =-I_{123}-I_{145}-I_{147}-I_{156}-I_{245}-I_{247}-I_{256}%
-I_{345}-I_{347}-I_{356}\nonumber\\
&  +I_{1245}+I_{1247}+I_{1256}+I_{1345}+I_{1347}+I_{1356}+I_{1456}%
+I_{1457}\nonumber\\
&  +I_{2345}+I_{2347}+I_{2356}+I_{2456}+I_{2457}+I_{3456}+I_{3457}\nonumber\\
&  -I_{12456}-I_{12457}-I_{13456}-I_{13457}-I_{23456}-I_{23457},
\end{align}
which is effectively a HEI in a 5-partite system.

By eliminating the region $A_{5}$, this HEI reduces to its joint form,
\begin{equation}
Q_{7}=-I_{123}-I_{167}-I_{267}-I_{367}+I_{1267}+I_{1367}+I_{2367}.
\label{Q7}
\end{equation}
By applying the permutation $(1,2,3,6,7)\rightarrow(2,3,4,1,5)$,  we can map the HEI $Q_{7}$ in eq.(\ref{Q7}) to $Q_{3}^{\left(  5\right)  }$ in eq.(\ref{Q53}), which has already been proved.

\section{Conclusion}

In this paper, we have developed a systematic strategy to prove Holographic
Entropy Inequalities (HEIs) in multipartite systems. This strategy integrates
several key conceptual and methodological ingredients.

First, we introduced the simplex basis. Any HEI in a fixed configuration can
be expressed in this basis and represented graphically by a circular diagram.
We analyzed the potential incompatibility between two connected RT surfaces and established the Compatibility Theorem, which led to the
definition of Compatible Completely Connected (CCC) configurations. We showed
that the circular diagram corresponding to a superbalanced HEI in a CCC
configuration is gapless and can be systematically simplified using the
clean-gap procedure. The resulting reduced diagram can then be directly
verified by cross inequalities.

Moreover, we demonstrated that the gapless structure of an HEI in a CCC
configuration allows it to be reduced to a joint form, which greatly
simplifies its verification.

To extend the analysis beyond CCC configurations, we introduced the critical
notion of a cut. By successively applying cuts to an HEI starting from a CCC
configuration, one can systematically explore all possible configurations. In
an $n$-partite system, there are in total
$\frac{n^{2}(n^{2}-1)}{12}$
cuts, organized into $(n-1)$ hierarchical levels. We proved the Cut Theorem,
which shows that cuts at higher levels induce cuts at lower levels, thereby
defining a dependency acyclic graph (DAG) structure among the cuts.

Finally, we established the Configuration Theorem, which states that if an HEI
holds in all CCC configurations, then it necessarily holds in every
configuration.

We applied our strategy to rigorously prove several example HEIs for systems
with $n = 3,4,5,6$, and $7$ parties. Having a systematic verification
framework in place opens the door to a systematic construction of new HEIs in
multipartite systems. At present, HEIs are fully understood only for
$n = 3,4,5$ and partially for $n = 6,7$. Constructing general HEIs for
higher-partite systems remains an extremely challenging problem. We plan to
address the systematic construction of general HEIs in multipartite systems in
future work.

\bibliographystyle{unsrt}
\bibliography{HEI}

@Article{0603001,
  author        = {Ryu, Shinsei and Takayanagi, Tadashi},
  journal       = {Phys. Rev. Lett.},
  title         = {{Holographic derivation of entanglement entropy from AdS/CFT}},
  year          = {2006},
  pages         = {181602},
  volume        = {96},
  archiveprefix = {arXiv},
  doi           = {10.1103/PhysRevLett.96.181602},
  eprint        = {hep-th/0603001},
  reportnumber  = {NSF-KITP-06-11},
}

@Article{0605073,
  author   = {Igor R. Klebanov and David Kutasov and Arvind Murugan},
  journal  = {Nuclear Physics B},
  title    = {Entanglement as a probe of confinement},
  year     = {2008},
  issn     = {0550-3213},
  number   = {1},
  pages    = {274-293},
  volume   = {796},
  doi      = {https://doi.org/10.1016/j.nuclphysb.2007.12.017},
  url      = {https://www.sciencedirect.com/science/article/pii/S0550321307009510},
}

@Article{0606184,
  author        = {Fursaev, Dmitri V.},
  journal       = {JHEP},
  title         = {{Proof of the holographic formula for entanglement entropy}},
  year          = {2006},
  pages         = {018},
  volume        = {09},
  archiveprefix = {arXiv},
  doi           = {10.1088/1126-6708/2006/09/018},
  eprint        = {hep-th/0606184},
}

@Article{0705.0016,
  author        = {Hubeny, Veronika E. and Rangamani, Mukund and Takayanagi, Tadashi},
  journal       = {JHEP},
  title         = {{A Covariant holographic entanglement entropy proposal}},
  year          = {2007},
  pages         = {062},
  volume        = {07},
  archiveprefix = {arXiv},
  doi           = {10.1088/1126-6708/2007/07/062},
  eprint        = {0705.0016},
  primaryclass  = {hep-th},
  reportnumber  = {DCPT-07-13, KUNS-2069},
}

@Article{1006.0047,
  author        = {Headrick, Matthew},
  journal       = {Phys. Rev. D},
  title         = {{Entanglement Renyi entropies in holographic theories}},
  year          = {2010},
  pages         = {126010},
  volume        = {82},
  archiveprefix = {arXiv},
  doi           = {10.1103/PhysRevD.82.126010},
  eprint        = {1006.0047},
  primaryclass  = {hep-th},
  reportnumber  = {BRX-TH-619},
}

@Article{1102.0440,
  author        = {Casini, Horacio and Huerta, Marina and Myers, Robert C.},
  journal       = {JHEP},
  title         = {{Towards a derivation of holographic entanglement entropy}},
  year          = {2011},
  pages         = {036},
  volume        = {05},
  archiveprefix = {arXiv},
  doi           = {10.1007/JHEP05(2011)036},
  eprint        = {1102.0440},
  primaryclass  = {hep-th},
}

@Article{1304.4926,
  author        = {Lewkowycz, Aitor and Maldacena, Juan},
  journal       = {JHEP},
  title         = {{Generalized gravitational entropy}},
  year          = {2013},
  pages         = {090},
  volume        = {08},
  archiveprefix = {arXiv},
  doi           = {10.1007/JHEP08(2013)090},
  eprint        = {1304.4926},
  primaryclass  = {hep-th},
}

@InBook{1609.01287,
  author    = {Rangamani, Mukund and Takayanagi, Tadashi},
  publisher = {Springer International Publishing},
  title     = {Holographic Entanglement Entropy},
  year      = {2016},
  booktitle = {title = "{Holographic Entanglement Entropy}", eprint = "1609.01287", archivePrefix = "arXiv", primaryClass = "hep-th", reportNumber = "YITP-16-106},
  date      = {2016},
  doi       = {10.1007/978-3-319-52573-0},
}

@Article{Lieb&Ruskai,
  author  = {Lieb, Elliott H. and Ruskai, Mary Beth},
  journal = {Advances in Mathematics},
  title   = {Some operator inequalities of the schwarz type},
  year    = {1974},
  issn    = {0001-8708},
  number  = {2},
  pages   = {269-273},
  volume  = {12},
  doi     = {https://doi.org/10.1016/S0001-8708(74)80004-6},
  url     = {https://www.sciencedirect.com/science/article/pii/S0001870874800046},
}

@Article{0704.3719,
  author        = {Headrick, Matthew and Takayanagi, Tadashi},
  journal       = {Phys. Rev. D},
  title         = {{A Holographic proof of the strong subadditivity of entanglement entropy}},
  year          = {2007},
  pages         = {106013},
  volume        = {76},
  archiveprefix = {arXiv},
  doi           = {10.1103/PhysRevD.76.106013},
  eprint        = {0704.3719},
  primaryclass  = {hep-th},
  reportnumber  = {SU-ITP-07-08, KUNS-2069, SU-ITP-07/08, KUNS-2069},
}

@Article{1211.3494,
  author        = {Wall, Aron C.},
  journal       = {Class. Quant. Grav.},
  title         = {{Maximin Surfaces, and the Strong Subadditivity of the Covariant Holographic Entanglement Entropy}},
  year          = {2014},
  number        = {22},
  pages         = {225007},
  volume        = {31},
  archiveprefix = {arXiv},
  doi           = {10.1088/0264-9381/31/22/225007},
  eprint        = {1211.3494},
  primaryclass  = {hep-th},
}

@Article{1107.2940,
  author        = {Hayden, Patrick and Headrick, Matthew and Maloney, Alexander},
  journal       = {Phys. Rev. D},
  title         = {{Holographic Mutual Information is Monogamous}},
  year          = {2013},
  number        = {4},
  pages         = {046003},
  volume        = {87},
  archiveprefix = {arXiv},
  doi           = {10.1103/PhysRevD.87.046003},
  eprint        = {1107.2940},
  primaryclass  = {hep-th},
  reportnumber  = {BRX-TH-638},
}

@Article{1505.07839,
  author        = {Bao, Ning and Nezami, Sepehr and Ooguri, Hirosi and Stoica, Bogdan and Sully, James and Walter, Michael},
  journal       = {JHEP},
  title         = {{The Holographic Entropy Cone}},
  year          = {2015},
  pages         = {130},
  volume        = {09},
  archiveprefix = {arXiv},
  doi           = {10.1007/JHEP09(2015)130},
  eprint        = {1505.07839},
  primaryclass  = {hep-th},
  reportnumber  = {CALT-TH-2015-020, IPMU15-0074, SLAC-PUB-16294, SU-ITP-15-08},
}

@Article{1612.02437,
  author        = {Walter, Michael and Gross, David and Eisert, Jens},
  title         = {{Multi-partite entanglement}},
  year          = {2016},
  month         = {12},
  archiveprefix = {arXiv},
  eprint        = {1612.02437},
  primaryclass  = {quant-ph},
}

@Article{1808.07871,
  author        = {Hubeny, Veronika E. and Rangamani, Mukund and Rota, Massimiliano},
  journal       = {Fortsch. Phys.},
  title         = {{Holographic entropy relations}},
  year          = {2018},
  number        = {11-12},
  pages         = {1800067},
  volume        = {66},
  archiveprefix = {arXiv},
  doi           = {10.1002/prop.201800067},
  eprint        = {1808.07871},
  primaryclass  = {hep-th},
}

@Article{1905.06985,
  author        = {He, Temple and Headrick, Matthew and Hubeny, Veronika E.},
  journal       = {JHEP},
  title         = {{Holographic Entropy Relations Repackaged}},
  year          = {2019},
  pages         = {118},
  volume        = {10},
  archiveprefix = {arXiv},
  doi           = {10.1007/JHEP10(2019)118},
  eprint        = {1905.06985},
  primaryclass  = {hep-th},
}

@Article{2003.03933,
  author        = {Guo, Wu-zhong},
  journal       = {JHEP},
  title         = {{Correlations in geometric states}},
  year          = {2020},
  pages         = {125},
  volume        = {08},
  archiveprefix = {arXiv},
  doi           = {10.1007/JHEP08(2020)125},
  eprint        = {2003.03933},
  primaryclass  = {hep-th},
}

@Article{2008.12430,
  author        = {Guo, Wu-zhong},
  journal       = {JHEP},
  title         = {{Entanglement spectrum of geometric states}},
  year          = {2021},
  pages         = {085},
  volume        = {02},
  archiveprefix = {arXiv},
  doi           = {10.1007/JHEP02(2021)085},
  eprint        = {2008.12430},
  primaryclass  = {hep-th},
}

@Article{1708.09393,
  author        = {Takayanagi, Tadashi and Umemoto, Koji},
  journal       = {Nature Phys.},
  title         = {{Entanglement of purification through holographic duality}},
  year          = {2018},
  number        = {6},
  pages         = {573--577},
  volume        = {14},
  archiveprefix = {arXiv},
  doi           = {10.1038/s41567-018-0075-2},
  eprint        = {1708.09393},
  primaryclass  = {hep-th},
  reportnumber  = {YITP-17-89, IPMU17-0115},
}

@Article{1709.07424,
  author        = {Nguyen, Phuc and Devakul, Trithep and Halbasch, Matthew G. and Zaletel, Michael P. and Swingle, Brian},
  journal       = {JHEP},
  title         = {{Entanglement of purification: from spin chains to holography}},
  year          = {2018},
  pages         = {098},
  volume        = {01},
  archiveprefix = {arXiv},
  doi           = {10.1007/JHEP01(2018)098},
  eprint        = {1709.07424},
  primaryclass  = {hep-th},
}

@Article{Araki&Lieb,
  author  = {Araki, Huzihiro and Lieb, Elliot H.},
  journal = {Communications in Mathematical Physics},
  title   = {Entropy inequalities},
  year    = {1970},
  issn    = {1432-0916},
  number  = {2},
  pages   = {160-170},
  volume  = {18},
  doi     = {https://doi.org/10.1007/BF01646092},
  url     = {https://link.springer.com/article/10.1007/bf01646092},
}

@Article{1802.09545,
  author        = {Bhattacharyya, Arpan and Takayanagi, Tadashi and Umemoto, Koji},
  journal       = {JHEP},
  title         = {{Entanglement of Purification in Free Scalar Field Theories}},
  year          = {2018},
  pages         = {132},
  volume        = {04},
  archiveprefix = {arXiv},
  doi           = {10.1007/JHEP04(2018)132},
  eprint        = {1802.09545},
  primaryclass  = {hep-th},
  reportnumber  = {YITP-18-12, IPMU18-0035},
}

@Article{1902.02369,
  author        = {Bhattacharyya, Arpan and Jahn, Alexander and Takayanagi, Tadashi and Umemoto, Koji},
  journal       = {Phys. Rev. Lett.},
  title         = {{Entanglement of Purification in Many Body Systems and Symmetry Breaking}},
  year          = {2019},
  number        = {20},
  pages         = {201601},
  volume        = {122},
  archiveprefix = {arXiv},
  doi           = {10.1103/PhysRevLett.122.201601},
  eprint        = {1902.02369},
  primaryclass  = {hep-th},
  reportnumber  = {YITP-19-05, IPMU19-0014},
}

@Article{2011.02790,
  author        = {Chou, Chia-Jui and Lin, Bo-Han and Wang, Bin and Yang, Yi},
  journal       = {JHEP},
  title         = {{Entanglement entropy inequalities in BCFT by holography}},
  year          = {2021},
  pages         = {154},
  volume        = {02},
  archiveprefix = {arXiv},
  doi           = {10.1007/JHEP02(2021)154},
  eprint        = {2011.02790},
  primaryclass  = {hep-th},
}

@Article{1812.08133,
  author        = {Hubeny, Veronika E. and Rangamani, Mukund and Rota, Massimiliano},
  journal       = {Fortsch. Phys.},
  title         = {{The holographic entropy arrangement}},
  year          = {2019},
  number        = {4},
  pages         = {1900011},
  volume        = {67},
  archiveprefix = {arXiv},
  doi           = {10.1002/prop.201900011},
  eprint        = {1812.08133},
  primaryclass  = {hep-th},
}

@Article{2002.04558,
  author        = {He, Temple and Hubeny, Veronika E. and Rangamani, Mukund},
  journal       = {JHEP},
  title         = {{Superbalance of Holographic Entropy Inequalities}},
  year          = {2020},
  pages         = {245},
  volume        = {07},
  archiveprefix = {arXiv},
  doi           = {10.1007/JHEP07(2020)245},
  eprint        = {2002.04558},
  primaryclass  = {hep-th},
}

@Article{2309.06296,
  author        = {Hern{\'a}ndez-Cuenca, Sergio and Hubeny, Veronika E. and Jia, Hewei Frederic},
  journal       = {JHEP},
  title         = {{Holographic entropy inequalities and multipartite entanglement}},
  year          = {2024},
  pages         = {238},
  volume        = {08},
  archiveprefix = {arXiv},
  doi           = {10.1007/JHEP08(2024)238},
  eprint        = {2309.06296},
  primaryclass  = {hep-th},
  reportnumber  = {MIT-CTP/5610},
}

\end{document}